\documentclass[a4paper,12pt]{article}
\usepackage{amssymb,amsmath,amsthm}
\usepackage{graphicx}
\usepackage{amsfonts}
\usepackage{latexsym}
\usepackage{color}
\usepackage[english]{babel}
\usepackage{hyperref}

\numberwithin{equation}{section}

\newtheorem{lemma}{Lemma}[section]
\newtheorem{theorem}[lemma]{Theorem}
\newtheorem{proposition}[lemma]{Proposition}
\newtheorem{corollary}[lemma]{Corollary}
\newtheorem{remark}[lemma]{Remark}

\newtheorem{definition}[lemma]{Definition}

\def\ts{{\tt s}}
\def\tT{{\tt T}}
\def\tr{{\tt r}}

\def\sgm{{\rm sgn}}

\def\cV{{\mathcal{V}}}

\def\cX{{\mathcal{X}}}
\def\cY{{\mathcal{Y}}}

\def\cL{{\mathcal{L}}}

\def\cU{{\mathcal{U}}}

\def\cP{{\mathcal{P}}}
\def\spazio#1{{\mathcal{B}}^{#1}}

\def\cT{{\mathcal{T}}}
\def\cTn{{\mathcal{T}_{N}}}
\def\Tn{{{T}_{N}}}

\def\Tr{{{T}}}
\def\uno{1}
\def\cB{{\mathcal{B}}}

\def\unp{\vskip10pt}

\def\poisson#1#2{\left\{#1;#2\right\}}
\def\poi#1#2{\left\{#1;#2\right\}}
\newcommand{\R}{\mathbb R}

\renewcommand{\O }{O }
\def\norma#1{\Vert#1\Vert}

\def\alla#1#2{#1^{(#2)}}
\def\xalla#1{X^{(#1)}}

\newcommand{\Kcs}{\tilde{\mathcal B}}
\def\ac{{\mathcal A}\kern-.7pt\ell\kern-.9pt\mathcal{S}}

\begin{document}


\title{{\bf Hamiltonian studies on counter-propagating water waves }}

\date{}


\author{ Dario Bambusi\footnote{Dipartimento di Matematica, Universit\`a degli Studi di Milano, Via Saldini 50, I-20133
Milano. 
 \textit{Email: } \texttt{dario.bambusi@unimi.it}}}

\maketitle

\begin{abstract}
We use a Hamiltonian normal form approach to study the dynamics of the
water wave problem in the small amplitude long wave regime (KdV
regime). If $\mu$ is the small parameter corresponding to the inverse
of the wave length, we show that the normal form at order $\mu^5$
consists of two decoupled equation, one describing right going waves
and the other describing left going waves. Performing a further non
Hamiltonian transformation we conjugate each of these equations to a
linear combination of the first three equations in the KdV hierarchy.
At order $\mu^7$ we find nontrivial terms coupling the two
counter-propagating waves.
\end{abstract}
\noindent





 
\section{Introduction}\label{intro}

In this paper we study the dynamics of the free surface of a fluid
which evolves under the influence of gravitation. The aim is to find
the effective equation governing the dynamics in the regime of small
amplitude and long wave. It is well known that, at the first
nontrivial order, the effective equation is the Kortweg de Vries
equation; more precisely, the dynamics is described by two KdV
equations \cite{WS}, one describing right going waves and the other
describing left going waves, moreover the two counter-propagating waves
do not interact, at least at the order of approximation controlled by
KdV.

Here starting from the so called Zakharov-Craig-Sulem Hamiltonian
approach to the water wave dynamics
\cite{Zak68,Cra,CraSul} we use Birkhoff
Normal form theory in order to attack the problem. As a first result
we get that the two decoupled KdV mentioned above are just
the Hamilton equations of the first order Birkhoff Normal Form of the
system. More generally, it turns out that {\it at any order}, the normal
form of the system consists just of two decoupled equations, one
describing right going waves and the other describing left going
waves. The problem is that, in order to put the system in normal form,
one has to construct a canonical transformation conjugating the
original Hamiltonian to its normal form, and the existence of such a
transformation is not ensured by any known general argument.
So, we investigate the existence of the normalizing transformation; we
prove that the transformation putting the
system in second order normal form exists, while we find an
obstruction to the existence of the transformation putting the system
in third order normal form.  To be
slightly more precise, let $\mu$ be a small parameter, and consider
an initial datum of size of order $\mu^2$ and wave length 
of order $\mu^{-1}$, then KdV is the normal form at order $\mu^3$; we
show that the system can be put in normal form at order $\mu^5$ and
that there is an obstruction to put the system in normal form at order
$\mu^7$.

So we stop our Hamiltonian construction at order 5 and analyze the
equation that we get. It turns out that this equation falls in a class
analyzed by Kodama (see \cite{Kod85, Kod87, Kod87a, HK09}), who showed
that there always exists a non-Hamiltonian transformation conjugating,
at order $\mu^5$, such an equation to a linear combination of the
first three equations of the KdV hierarchy. Remarkably enough, this is
not true at order $\mu^7$. Thus we apply Kodama's result getting that,
up to order $\mu^7$, counterpropagating waves are described by {\it two
decoupled non iteracting equations, each of which is an integrable
equation} which is a linear combination of the first three equations in
the KdV hierarchy.

We emphasize that the idea of using the Hamiltonian approach to show
the appearance of KdV in water wave theory appeared in \cite{Cra},
where Craig and Groves made an expansion of the Hamiltonian in powers
of the parameter $\mu$ (the one we just introduced) and then studied
the first terms of the so obtained Hamiltonian in order to find the
effective equations.  A fundamental step in their procedure (a step
which plays a crucial role also in the present paper) consists in
parametrizing the surface of the fluid using suitable functions
$r(y,t)$, $s(y,t)$, where $t$ is a rescaled time variable and $y$ is a
rescaled space variable. Then the equations of motion of the
unperturbed system turn out to be given simply by
\begin{equation}
  \label{char}
\frac{\partial r}{\partial t}=-\frac{\partial r}{\partial y}\ ,\quad
\frac{\partial s}{\partial t}=\frac{\partial s}{\partial y}\ ,
\end{equation}
whose solution is of course a right going wave non interacting with a
left going wave. For this reason we will call such functions
characteristic variables. Then the main remark of \cite{Cra}
(concerning KdV) is that, if one restricts the Hamiltonian to the
submanifold $s=0$, then the Hamiltonian turns out to coincide with the
Hamiltonian of the KdV equation. The same is true for the Poisson
tensor so that, in this submanifold, the equation of motion coincide
with the KdV equation.  However, with this procedure one does not see
the appearance of the second KdV equation, and furthermore one has the
problem that the manifold $s=0$ is not invariant under the
dynamics. Here normal form theory comes into play: indeed, using the
characteristic variables, it is very easy to compute the first order
normal form of the system and to get that it consists just of a couple
of decoupled KdV equations. This method was already used in the
context of the FPU problem in \cite{BP06} and a similar point of view
was also used in \cite{BCP02} in order to deduce the NLS equation as a
normal form for the Klein Gordon equation.
Now, once one has computed the first term of the normal form, it is
very natural to try to iterate the procedure. In this way we get our
Hamiltonian result, and then, as anticipated above we perform Kodama's
transformation in order to reduce our equations to a couple of decoupled
integrable equations.

\vskip10pt

We now recall a few results on the deduction of modulation equations
for the water wave problem. First, it is by now quite standard to
obtain KdV as an equation describing unidirectional waves; KdV$_5$ has
also been deduced as an higher order approximation for such
unidirectional waves (see e.g. \cite{DGH03}). For the case of more
general initial data, giving rise to counterpropagating waves, we
recall that the corrections of order $\mu^5$ to the modulation
equation were studied in \cite{Wri05}, where the author
obtained that the first correction to the KdV equation contains terms
which fulfill a linear time dependent equation plus terms in which an
interaction of the counter-propagating waves is actually present. We
emphasize that this description is compatible with the description
that we get here. In particular the interaction between the
counter-propagating waves is a product of the coordinate
transformation that we use to put the system in normal form.  A
remarkable fact that our description yields concerning the interaction
of counter-propagating waves is that the effects of interaction between the two
waves disappear after the interaction, so that, if two spatially
localized waves cross, then after the interaction they should return
to the original shape, at least at the considered order of
approximation. 

We also recall that the modulation equation that describes the
solutions can depend on the kind of initial data
that one considers, in particular on the decay at infinity of the
data. An interesting discussion of this phenomenon can be found in
\cite{BCL05} (see also \cite{Lan19}).

A final consideration pertains the dynamics of the water waves in the
complete model: after proving that a solution of the normal form
equation fulfills the equations of the water wave problem up to an
error of order $\mu^7$, we apply Theorem 4.18 of \cite{Lannes} to
prove that the solution of the water wave problem remains $O(\mu^4)$
close to the solution of the normal form equation for a time of order
$\mu^{-3}$. We emphasize that the result applies to {\it all} initial
data for the water wave problem which are of class $W^{\ts,1}$ with
$\ts$ large enough. However such a time scale is not very satisfactory, since
the time over which the dynamics of the fifth order normal form
becomes visible is $\mu^{-5}$. It would be ineresting to try to apply
the technique recently introduced in \cite{BD18} (see also
\cite{BFP18}) in order to reach such a longer time scale.

\unp

From a technical point of view, the proof of our result requires some
nontrivial steps. First one has to develop a normal form technique in the
case where the unperturbed system is essentially a transport equation
on $\R$. Actually some averaging techniques adapted to this situation
were already developed in \cite{BCP02}. Here, due to the particular
structure of the water wave problem, we find that such techniques are
particularly effective, and in particular we find a general algorithm to
solve the so called homological equation.

The main difficulty is related to the fact that, in Hamiltonian
perturbation theory, the transformation conjugating the system to its
normal form is typically generated as the flow of some auxiliary
Hamiltonian system. However, it turns out that the auxiliary
Hamiltonian system one finds does not generate a flow (it is very
similar an inverse heat equation). In Sect.\ref{BNF} we develop a
technique allowing to put the system in normal form in the case of
vector fields not generating a flow. The idea is to approximate the
flow through its truncated expansion in the small parameter involved
in the construction. The nontrivial point is that the so obtained
transformation is not canonical, but only approximately canonical,
thus one has to show that it can actually be used to normalize the
system at the wanted order of approximation. We mention that an
alternative technique that one could try to use in order to normalize
the systems is that introduced in \cite{Bam05} (also used in
\cite{BP06}), which is based on the use of Galerkin truncations. Maybe
this would work also here, but this is not clear due to the
difficulty of defining in a suitable way the operator $\partial^{-1}$
in Fourier transform.

The paper is organized as follows In Sect. \ref{s.main} we give our
main result; in Sect. \ref{prelimi} we prepare the Hamiltonian of the
water wave problem for the application of the normal form
procedure. In particular this section reproduces the procedure by
Craig and Groves in order to deduce KdV. In Sect. \ref{BNF} we develop
an abstract framework for Hamiltonian normal form in the case of
vector fields that do not generate a flow.  In Sect. \ref{NF} we
develop the tools needed to solve the so called homological equation
in the case of the water wave problem and we prove our result on
Hamiltonian normal form. We also show the obstruction that one finds
when trying to put the system in normal form at order
$\mu^7$. Finally, in Sect. \ref{kodama}, we recall Kodama's
transformation and conclude the proof of our main theorem. 

\vskip1 truecm

\noindent {\it This paper is dedicated to the memory of Walter Craig,}
he was a good friend and from a scientific point of view he had a
great influence on my work. It was always a great pleasure to meet
Walter and to spend time with him discussing about science or doing
sport and tourism. I miss his great humanity and his enthusiasm.

\vskip1truecm
\noindent {\it Acknowledgments.} Part of the material presented in
this paper is the content of some lectures that I gave more than 10
years ago in order to prepare a visit by Walter Craig. I thank all the
people who attended such lectures and contributed with their comments
to improve the material, in particular Antonio Ponno with whom I had a
lot of discussions on the subject. The discussions with Antonio Ponno
were also the key to the understanding of the relevance of Kodama's
work in the present context. I also would like to thank Doug Wright
and David Lannes who gave me some relevant feedbacks on higher order
corrections to KdV and on some technical issues. Finally I warmly
thank the two referees of the paper whose comments allowed to greatly
improve the paper.

\section{Main result}\label{s.main}

Consider an ideal fluid occupying, at rest, the domain
$$
\Omega_0:=\left\{(x,z)\in\Re^2\ :\ -h<z<0\right\}\ ,
$$ we study the evolution of the free surface under the action of
gravity, in the irrotational regime. Thus, given a function $\eta(x)$,
we define the domain
\begin{equation}
  \label{dom}
\Omega_\eta:=\left\{(x,z)\in\Re^2\ :\ -h<z<\eta(x)\right\}\ .
\end{equation}
and introduce the velocity potential $\phi$,
which is related to the velocity of the fluid by $u=\nabla \phi$. It
is well known that the problem admits a Hamiltonian
formulation \cite{Zak68,Cra,CraSul}, the conjugated canonical variables being the wave profile
$\eta$ and the trace of the velocity potential at the free surface,
namely
\begin{equation}
  \label{psi}
\psi(x):=\phi(x,\eta(x))\ .
\end{equation}

For the moment, just to fix ideas we work in the phase space of
functions $z\equiv(\eta,\psi)$ of Schwartz class, later we will work
in a more general setting. 
We endow the phase space by the $L^2$ scalar product,
namely
$$
\langle
z;z'\rangle=\langle(\eta,\psi);(\eta',\psi')\rangle:=\langle\eta;\eta'\rangle_{L^2}+\langle\psi;\psi'\rangle_{L^2}\ .
$$
and by the Poisson tensor
\begin{equation}
  \label{poisson}
J(\eta,\psi):=(-\psi,\eta)\ ,
\end{equation}
so that, given a Hamiltonian function $H=H(z)$, and defining its $L^2$
gradient (which is defined by $dH(z)h=\langle\nabla H(z);h\rangle$) the
Hamilton equations are given by 
\begin{equation}
  \label{hameq}
  \dot z=J\nabla H(z)\iff \left\{
  \begin{matrix}
    \dot \eta=\nabla_{\psi}H(\eta,\psi)
    \\
    \dot \psi=-\nabla_{\eta}H(\eta,\psi)
  \end{matrix}
  \right. \ .
\end{equation}

The Hamiltonian of the water wave problem is given by 
\begin{equation}
  \label{WW}
H(\eta,\psi)=\int \left(\frac{1}{2}g\eta^2+\frac{1}{2}\psi
G(\eta)\psi\right)dx
\end{equation}
and $G$ is the Dirichlet Neumann operator (see Definition \ref{gn}).

We will look for solutions of the form
\begin{equation}
  \label{vecchie1.i}
\eta(x)=\mu^2h^3\sqrt2\ \tilde
\eta(\mu x) \ ,\qquad \psi(x)=\mu \sqrt{2gh}h^2\ \tilde
\psi(\mu x) \ ,\quad \mu\ll 1\ ,
\end{equation}
where the factors depending on $g$ and $h$ have been inserted for
future convenience.
In terms of the variables $\tilde \eta$ and $\tilde \psi$ the system
is still Hamiltonian with a scaled Hamiltonian (see
Subsect. \ref{scaling}), which takes the form
\begin{equation}
  \label{hscalata}
\mu\sqrt{gh}  H_{WW}(\tilde \eta,\tilde \psi)\ ,
\end{equation}
with a suitable smooth $H_{WW}$. So, dividing by $\mu\sqrt{gh}$, which is equivalent to pass to the scaled
time
\begin{equation}
  \label{ttilde}
\tilde t:=\frac{t}{\mu\sqrt{gh}}\ ,
\end{equation}
one is reduced to the Hamiltonian system with Hamiltonian $
H_{WW}$. Expanding in $\mu$ it takes the form (see Subsect. \ref{expa})
\begin{equation}
  \label{H.riscl.i}
H_{WW}=H_0+\epsilon H_1+\epsilon^2 H_2+...
\end{equation}
where $\epsilon:=(h\mu)^2$ 
\begin{align}
  \label{vera.i}
H_0=\int
\frac{ \eta^2+ \psi_y^2 }{2}dy
\end{align}
and the expressions of the higher order terms are not relevant for the
moment. Here we also omitted the tildes.

Then, following \cite{Cra}, it is convenient to introduce the
characteristic variables 
\begin{equation}
  \label{rs}
r=\frac{\eta+\psi_y}{\sqrt2}\ ,\quad s=\frac{\eta-\psi_y}{\sqrt2} \ ,
\end{equation}
which transform the Poisson tensor essentially in the Poisson tensor
of the KdV equation (see Remark \ref{B} and
Subsect. \ref{char.s}). Precisely, the Hamilton equations of a
Hamiltonian $H(r,s)$ turn out to be given by
\begin{equation}
  \label{rsH}
\dot r=-\partial_y\nabla _rH\ , \quad \dot s=\partial_y\nabla _sH \ .
\end{equation}
In particular one has that $H_0$ takes the form 
\begin{align}
  \label{h0rs.i}
H_0&=\int\frac{r^2+s^2}{2}dy\ ,
\end{align}
whose equations of motion are given by \eqref{char}.

We remark that, \eqref{rs} is just a change of variables, so that, if
a solution is written in terms of the variables $r=r(y,t)$ and
$s=s(y,t)$, then one
can go back to the rescaled physical variables 
\begin{align}
  \label{phisiche1}
\eta(y,t)&:=\frac{1}{\sqrt2}\left[
r\Big(y,t
\big)+s\big(y,t \Big)  \right]\ ,
\\
\label{phisiche2}
\psi_y(y,t)&:=\frac{1}{\sqrt2}\left[
r\Big(y,t
\big)-s\big(y, t\Big)  \right]\ 
\end{align}
(originally denoted by $\tilde \eta$, $\tilde \psi$) in order to get
the wave profile and the trace of the velocity potential.  We remark that
the integration constant allowing to pass from $\psi_y$ to $\psi$ is
invariant with respect to the dynamics, so it is irrelevant in the
following.

In particular it is possible to rewrite the Hamiltonian $H_{WW}$ (c.f.
\eqref{H.riscl.i}) in terms of the variables $(r,s)$.  \emph{We will
  still say that this is the Hamiltonian of the water wave problem (in
  the variables $(r,s)$)}.

\begin{definition}
  \label{physical}
{In the following, given a couple of function
  $r(y,t)$, $s(y,t)$, we say that
\begin{equation}
  \label{back}
z^p(y,t):=(\eta(y,t),\psi(y,t))\ ,
\end{equation}
with $\eta,\psi$ given by \eqref{phisiche1} and \eqref{phisiche2} is
called 
\emph{the corresponding function in scaled physical variables.}
}\end{definition}

As anticipated in the introduction, our goal is to put the system in
normal form at second order.

Before stating the main result we still need a few
preliminaries.

First we recall that the KdV hierarchy consists of a sequence of
Hamiltonian systems with Hamiltonians $K_0,K_1,K_2...$, each of which
is integrable and which pairwise commute, so that, in some sense, they
form a complete set of integral of motion. Given an abstract function
$u=u(y)$, the first three Hamiltonians are given explicitely
(with a suitable choice of a normalization parameter) by
\begin{align}
  \label{kdv0}
  K_0(u)&=\frac{1}{2}\int_{\R}u^2dy\ ,
  \\
  \label{kdv1}
  K_1(u)&=\int_{\R}\left(-\frac{1}{12}u^2_y+\frac{1}{3}u^3
  \right)dy\ ,
  \\
  \label{kdv2}
  K_2(u)&=\int_{\R}\left(\frac{1}{2}u^2_{yy}-\frac{5}{2}u^2_xu+\frac{5}{8}u^4
  \right)dy\ ,
\end{align}
and the corresponding Hamilton vector fields are
$$
\frac{du}{dt}=-\partial_y\nabla K_j(u)\ ,\quad j=0,1,2\ .
$$
Consider also the following Hamiltonian
\begin{align}
  \label{hanf}
  H_{NF}(r,s):=&K_0(r)+\epsilon K_1(r)+\epsilon^2c_2K_2(r)
  \\
  \label{hanf.1}
  +& K_0(s)+\epsilon K_1(s)+\epsilon^2c_2K_2(s)\ ,
\end{align}
(with an arbitray $c_2\in\R$) and remark that its Hamilton equations
are two decoupled equations, one for $r$ and one for $s$. Each of
these equations is integrable and one passes from one to the other
just by inverting the space, namely by the transformation $y\to -y$.

Then, in order to precisely specify the properties of the transformation
$T_\epsilon$ used to conjugate to the final normal form, we need to
define the operator $\partial^{-1}$
by 
\begin{equation}
  \label{dmeno.1.i}
(\partial^{-1}u)(y):=\frac{1}{2}\left[\int_{-\infty}^yu(y_1)dy_1
    -\int_y^{+\infty}u(y_1)dy_1\right]\ .
\end{equation}

Finally, we define the precise phase space we are going to use: we will
work in the scale of Banach spaces $\spazio {\ts}:=W^{\ts ,1}\times
W^{\ts,1}\ni (r,s)\equiv z$ where $W^{\ts,1}$ is the Sobolev space of
the $L^1$ functions which have weak derivatives of order $\ts$ of
class $L^1$. We consider the case $\ts\gg1$. We will denote by
$B_1^{\ts}\subset\spazio {\ts}$ the ball of radius 1 centered at the
origin. We will also consider the Sobolev spaces $W^{\ts ,2}$ based on
$L^2$. 

\begin{theorem}
  \label{main.0}
For any $\ts'$ there exists $\epsilon_*>0$ and $\ts,\ts''$, s.t., if $0<\epsilon<\epsilon_*$, then
there exists a map $T_\epsilon:B_1^{\ts}\to W^{\ts'',1}\times W^{\ts'',1}$, with
the following properties
\begin{itemize}
\item[(i)] $T_\epsilon(r,s)-(r,s)$ is a polynomial in
  $\partial^{k}r,\partial^k s$, $k=-1,...,5$,
\item[(ii)] $\sup_{{(r,s)\in B_1^{\ts}}}\norma{T_\epsilon(r,s)-(r,s)}_{W^{\ts'',1}\times W^{\ts'',1}
}\leq C \epsilon$,
\item[(iii)] Let $I_\epsilon$ be an interval containing the origin and $z(.)=(r(.),s(.))\in C^1(I_{\epsilon};B_1^{\ts})$ be a solution
  of the Hamiltonian system \eqref{hanf}, \eqref{hanf.1}, with
  $c_2=\frac{299}{389}$,  
define
  \begin{equation}
  \label{trans}
z_a\equiv(r_a,s_a):=T_{\epsilon}(r,s)\ .
\end{equation}
Then there exists $R\in C^1(I_\epsilon,W^{\ts',2}\times
W^{\ts',2})$ s.t. one has
  \begin{equation}
    \label{main.diff}
\dot z_a(t)=J\nabla H_{WW}(z_a(t)) + \epsilon^3 R(t)\ ,\quad \forall t\in
I_\epsilon\ ,
  \end{equation}
  where $H_{WW}$ is the Hamiltonian \eqref{H.riscl.i} of the water wave problem
  rewritten in the variables $(r,s)$.
\end{itemize}
\end{theorem}

\begin{remark}
  \label{explit}
The explicit form of the transformation $T_\epsilon$ is not computed
explicitly in the paper, but it can be extracted by slghtly developing
the computations of Sect. \ref{NF}. It is given by $T_1\circ T_2\circ
T_K$, with  $T_1$, $T_2$ and $T_K$ given by \eqref{t1.eq},
\eqref{t2.eq} and \eqref{kod.2} respectively.
\end{remark}

By applying Theorem 4.18 of \cite{Lannes} one immediately gets the
following result (here the smoothness indexes have a value different
from those of Theorem \ref{main.0}) which gives some dynamical
information on all solutions with smooth enough integrable initial
data.

\begin{corollary}
  \label{main}
For any $\ts'$ there exist $\ts$, $\epsilon_*>0$ and $\tT>0$ s.t., if
$\epsilon<\epsilon_*$ then the
following holds true. Consider the Cauchy problem for the water wave
problem $H_{WW}$ with initial datum $(\eta_0,\psi_0)$ fulfilling
\begin{equation}
  \label{ini.c}
\norma{(\eta_0,\psi_0)}_{W^{\ts,1}\times
  W^{\ts+1,1} }\leq 1\ 
\end{equation}
and denote by $z^p(t)$ the corresponding solution (of course still in
the scaled physical variables $(\eta,\psi)$). Then there exists a
solution $(r(t),s(t))$ of the Hamiltonian system  \eqref{hanf},
\eqref{hanf.1} with the following property:
denote by $z_a$ the function defined by \eqref{trans}, and by $z^p_a$
the correspoding solution in scaled physical variables, then one has 
\begin{equation}
    \label{main.est}
\sup_{|t|\leq \tT/\epsilon}\norma{z^{p}_a(t)-z^p(t)}_{W^{\ts',2}\times
  W^{\ts',2}}\leq C\epsilon^2\ .
  \end{equation}
\end{corollary}
From this Corollary, it is possible to go back to the {\it original
  non scaled} variables. For example, exploiting the embedding of
$W^{\ts',2}\subset L^{\infty}$ (provided $\ts'>1/2$), one can get the following estimate on the
profile of the wave:
  \begin{equation}
    \label{main.est1}
\sup_{|t|\leq \tT/\mu^3\sqrt{gh}}\norma{\eta(t)-\eta_a(t)}_{L^{\infty}}\leq C\mu^6\ .
  \end{equation}

We remark that in these results we never tried to get optimality
concerning the regularity loss related to our procedure.

A more serious problem with our deduction is that we deal with
soultions of the normal form equations which belong to the Sobolev
spaces based on $L^1$. It is not clear if this can be avoided. This
plays a role in the definition of $\partial^{-1}$. One could try to work in spaces of square integrable
functions in which there is a better theory of existence and
uniqueness for the equations of the KdV hierarchy, but this requires
some nontrivial work. Probably one could work with some regularized
version of the operator $\partial^{-1}$ defined in \eqref{dmeno.1.i}.

Finally we remark that, as anticipated in the introduction, one would
like to get results of correspondence between the approximate solution
and the true solutions over longer time scales, but this is beyond the
schope of this paper and is left for future work.

\section{Preliminaries: scaling, expansions and characteristic variables}
\label{prelimi}

\subsection{Canonical transformations}\label{canonical}

In this subsection we recall a few basic facts of Hamiltonian
mechanics

Consider a change of variables $z:=T(\zeta)$. Exploiting the formula
\begin{equation}
  \label{nablecirc}
\nabla(H\circ T)(\zeta)=[dT(\zeta)]^*(\nabla H)(T(\zeta))\ ,
\end{equation}
which is true for any smooth function $H$, one immediately sees that if $\zeta(t)$ fulfills the Hamiltonian equations
  \begin{equation}
    \label{hamaet}
\dot\zeta=J\nabla(H\circ T)(\zeta) \ ,
  \end{equation}
  then, $z(t)=T(\zeta(t))$  fulfills
  \begin{equation}
    \label{tra.ham}
\dot z=dT(\zeta)\dot \zeta=dT(\zeta)J[dT(\zeta)]^*(\nabla
H)(T(\zeta))\ .
  \end{equation}
  Viceversa, if $z$ fulfills
  $$\dot z=J\nabla H(z)\ ,$$
  then one has
  $$
\dot\zeta=[dT^{-1}(T(\zeta))]J[dT^{-1}(T(\zeta))]^*\nabla (H\circ
T)(\zeta)\ ,
  $$
where the star denotes the adjoint with respect to the $L^2$ scalar
product. 

With this remark one can characterize the coordinate transformations
leaving invariant the Hamiltonian formalism.

\begin{definition}
  \label{canonical}
A coordinate transformation $z=T(\zeta)$ is said to be canonical if
it transforms the Hamilton equation of any Hamiltonian $H$ into the
Hamilton equations of $H\circ T$.
\end{definition}

As a consequence of eq. \eqref{tra.ham} we have the following
proposition. 

\begin{proposition}
  \label{can.65}
  A coordinate transformtion is canonical if and only if it fulfills
  \begin{equation}
    \label{can.68}
dT(\zeta)J[dT(\zeta)]^*=J\ .
  \end{equation}
\end{proposition}
Then it is immediate to see that \eqref{can.68} is equivalent to 
$$
[dT^{-1}(z)]J[dT^{-1}(z)]^*=J\ .
$$

\subsection{Hamiltonian scalings}\label{scaling}

The main tool in order to perform the scaling
at a Hamiltonian level is the following remark which is also needed in
order to compute how the Poisson tensor changes when introducing the
characteristic variables. 

\begin{remark}
  \label{B}
A linear change of variables $
z=B\zeta $, transforms the Hamilton equations of $H$ into the equations
$\dot{\zeta }=\tilde J\nabla\widehat{ H}(\zeta )$, where $\widehat {
H}(\zeta ):=H(B\zeta )$ and $\tilde J:=B^{-1}JB^{-*}$, and $B^{-*}$
is the adjoint (with respect to the $L^2$ metric) of the inverse of $B$. 
\end{remark}
In the particular case of the linear change of coordinates given
by
\begin{equation}
  \label{scaling.1}
  z=B\zeta \qquad\iff\qquad \left\{\begin{matrix}
  \eta(x)=\epsilon_1\tilde \eta(\mu x)
  \\
  \psi(x)=\epsilon_2\tilde \psi(\mu x)
\end{matrix}\right.
\end{equation}
 One has the following Lemma
\begin{lemma}
  \label{lintra}
The transformation \eqref{scaling.1} transforms the Hamilton equations
of $H$ into the Hamilton equations 
\begin{equation}
  \label{tildeh}
\tilde H(\zeta ):=\frac{\mu}{\epsilon_1\epsilon_2} H(B\zeta )\ .
\end{equation}
\end{lemma}
\proof We just compute $B^{-1}$, $B^{-*}$ and $B^{-1}JB^{-*}$. First,
one has that $B^{-1}$ is given by
$$
[B^{-1}(\eta,\psi)](y)=\left(\frac{1}{\epsilon_1}\eta\left(\frac{y}{\mu}\right)
,\frac{1}{\epsilon_2}\psi\left(\frac{y}{\mu}\right) \right) \ , 
$$
from which one can compute its adjoint. Of course it enough to
consider one of the components of the vector $z$. We have
\begin{align*}
\langle \eta';B^{-1}\eta\rangle=\int \eta'(y)
\frac{1}{\epsilon_1}\eta\left(\frac{y}{\mu}\right)dy =\int 
\mu \eta'(y)
\frac{1}{\epsilon_1}\eta\left(\frac{y}{\mu}\right)d\frac{y}{\mu}
\\
=\mu\int  \eta(\mu x)
\frac{1}{\epsilon_1}\eta\left({x}\right)dx =  \langle B^{-*}\eta';\eta\rangle
\end{align*}
so that  we have
$$
[B^{-*}(\eta,\psi)](x)=\left(\frac{\mu}{\epsilon_1}\eta\left({\mu x}\right)
,\frac{\mu}{\epsilon_2}\psi\left(\mu{x}\right) \right) \ .
$$
It follows that
\begin{align}
[B^{-1}JB^{-*}(\eta,\psi)](y)=B^{-1}J\left(\frac{\mu}{\epsilon_1}\eta\left({\mu x}\right)
,\frac{\mu}{\epsilon_2}\psi\left(\mu{x}\right) \right)
\\
=B \left(\frac{\mu}{\epsilon_2}\psi\left({\mu x}\right)
,\frac{\mu}{\epsilon_1}\eta\left(\mu{x}\right) \right) =\left(
\frac{\mu}{\epsilon_1\epsilon_2}\psi(y),-
\frac{\mu}{\epsilon_1\epsilon_2}\eta(y)\right)
=\frac{\mu}{\epsilon_1\epsilon_2} J(\eta,\psi).
\end{align}
Thus the Hamilton equations of $H$ are transformed into $\dot \zeta=\frac{\mu}{\epsilon_1\epsilon_2} J \nabla \widehat{H}(\zeta )=J\nabla
\frac{\mu}{\epsilon_1\epsilon_2}\widehat{H}=J\nabla\tilde H $\qed 

\subsection{Expansion of the Hamiltonian}\label{expa}

In this subsection and in Subsect. \ref{char.s}, essentially we
repeat with minor changes the procedure developed in \cite{Cra} in
order to show the appearance of KdV in the water wave problem.

For the sake of completeness, we start by recalling the definition of
the Dirichlet-Neumann operator, then we will recall its expansion,
which was computed in \cite{CraSul}. 

\begin{definition}
  \label{gn}
Given a function $\psi(x)$, consider the boundary value problem 
\begin{align}
  \label{dn.1}
  \Delta\phi&=0 \qquad\qquad\qquad (x,z)\in\Omega_{\eta}
  \\
  \label{dn.2}
  \phi_z\Big|_{z=-h}&=0
  \\
   \label{dn.2.1}
\lim_{x\to\infty}\phi&=0
 \\
  \label{dn.3}
 \phi\Big|_{z=\eta(x)}&=\psi\qquad\qquad \quad ,
\end{align}
and let $\phi$ be its solution. 
Then the linear operator $G(\eta)$
defined by 
\begin{equation}
  \label{dn.4}
G(\eta)\psi=\sqrt{1+\eta_x^2} \partial_n\phi\big|_{
z=\eta(x)}\equiv \left(\phi_z-\eta_x\phi_x\right)\big|_{
z=\eta(x,y)}\ \ 
\end{equation}
is called the \emph{Dirichlet Neumann operator}, where $\partial_n$
is the derivative in the direction normal to $z=\eta(x)$.
\end{definition}

Formally, it is well known \cite{CraSul} that the Dirichlet
Neumann operator has a Taylor expansion of the form
$G(\eta)\simeq \sum_{j\geq 0}G^{(j)}(\eta)$ with $G^{(j)}(\eta)$ homogeneous
of degree $j$ in $\eta$. One has
\begin{align}
  \label{expaG.0}
  G^{(0)}&=D\tanh(hD)\ ,
  \\
  \label{expaG.1}
  G^{(1)}&=D\eta D- G^{(0)}\eta G^{(0)}
  \\   \label{expaG.2}
G^{(2)}&=-\frac{1}{2} \left( D^2\eta^2 G^{(0)}+G^{(0)}\eta^2 D^2-2
G^{(0)}\eta  G^{(0)}\eta G^{(0)}\right)
\end{align}
where we used the standard notation $D:=-i\partial_x$. 

Substituting \eqref{scaling.1} in \eqref{expaG.0}, denoting as above, by $y:=\mu
x$, and $\partial_y$ the
corresponding partial derivative and $D_y:=-i\partial_y$, one gets
\begin{align}
\nonumber
G^{(0)}&=\mu^2hD_y^2-\frac{1}{3}\mu^4h^3D^4_y+\frac{2}{15} \mu^6h^5D^6+\O(\mu^8)
\\
\label{Gex}
&=-\mu^2h\partial_y^2-\frac{1}{3}\mu^4h^3\partial^4_y-\frac{2}{15}
\mu^6h^5\partial_y^6 +\O(\mu^8)
\\  \nonumber
G^{(1)}&=\mu^2\epsilon_1D_y\eta D_y-\mu^4\epsilon_1 h^2 D_y^2\eta
D_y^2+\O(\epsilon_1\mu^6)
\\
\label{G1}
& =-\mu^2\epsilon_1\partial_y\eta \partial_y-\mu^4\epsilon_1
h^2 \partial_y^2\eta \partial_y^2+\O(\epsilon_1\mu^6)
\\
\nonumber
G^{(2)}&=\O(\epsilon_1^2\mu^4)\ .
\end{align}

Inserting in the Hamiltonian the scaling \eqref{scaling.1}, and the
expansions \eqref{Gex} and \eqref{G1} and taking advantage of Lemma
\ref{lintra}, one gets that the Hamiltonian for the scaled variables
becomes $H=H_0+H_1+H_2+h.o.t.$ with
\begin{align}
  \label{H0}
H_0&:=\frac{1}{2}\int
\left(\frac{\epsilon_1}{\epsilon_2}g \eta^2+\frac{\epsilon_2}{\epsilon_1}
\mu^2 h  \psi_y^2 \right)dy\ ,
\\
\label{H1}
H_1&:=\frac{1}{2}\int\left(-\frac{\epsilon_2}{\epsilon_1}\frac{1}{3}\mu^4
h^3 \psi_{yy}^2+\epsilon_2\mu^2 \eta \psi_y^2\right)dy\ ,
\\
\label{H2}
H_2&:=\frac{1}{2}\frac{\epsilon_2}{\epsilon_1}\int\left(\frac{2}{15}\mu^6
h^5 \psi_{yyy}^2-\mu^4\epsilon_1 h^2 \eta\psi_{yy}^2 \right)dy\ , 
\end{align}
where we omitted the tildes
(remark that, as a difference with the notation of Sect. \ref{s.main}, the
small parameters are here included in $H_j$. We will come back to the
original notation at the
end of this subsection). 
The choice
$$
\frac{\epsilon_1}{\epsilon_2}g=\frac{\epsilon_2}{\epsilon_1}\mu^2h
$$ makes the two terms of $H_0$ of equal order of magnitude, and gives
it the form
\begin{equation}
  \label{H0.1}
H_0:=\mu\sqrt{gh}\frac{1}{2}\int
\left(\eta^2+ \psi_y^2 \right)dy\ ;
\end{equation}
The choice $\epsilon_2=\mu h^2\sqrt{2gh}$, which implies
$\epsilon_1=\sqrt2\mu^2 h^3
$, also implies that the two terms of
$H_1$ have the same order of magnitude ($\sqrt2$ has been inserted for
future convenience).
Remark in particular that the relationship \eqref{scaling.1} turns out to take
the form \eqref{vecchie1.i}.

Inserting in the Hamiltonian one gets
\begin{align}
   \label{H111}
H_1:=\mu^3\sqrt{gh} h^2 \frac{1}{2}\int\left(-\frac{1}{3}
\psi_{yy}^2+
\sqrt2\eta \psi_y^2\right)dy
\\
\label{H211}
H_2:=\mu^5 \sqrt{gh}h^4\frac{1}{2}\int \left(\frac{2}{15}
\psi_{yyy}^2 -\sqrt2\eta\psi_{yy}^2\right) dy\ .
\end{align}
Finally passing to the scaled time $\tilde t$ (cf \eqref{ttilde}) and
separating the small parameter from $H_j$, the
Hamiltonian takes the form
\begin{equation}
  \label{H.riscl}
H_{WW}= H_0+\epsilon H_1+\epsilon^2 H_2+\O(\epsilon^3)\ , 
\end{equation}
with $\epsilon:=(h\mu)^2$ and
\begin{align}
  \label{vera}
H_0&=\int
\frac{ \eta^2+ \psi_y^2 }{2}dy
\\
H_1&=\frac{1}{2}\int\left(-\frac{1}{3}\psi_{yy}^2+\sqrt2\eta\psi_y^2\right)dy
\\
H_2&=\frac{1}{2}\int\left(\frac{2}{15}\psi_{yyy}^2-\sqrt2\eta\psi_{yy}^2\right)dy
\end{align}

More precisely,
we have the following result

\begin{proposition}
  \label{approh}
Consider the Hamiltonian \eqref{WW} and introduce the scaled variables
\eqref{scaling.1}. Let $H_{WW}$ be the scaled Hamiltonian of the
water wave problem in scaled variables, then, for any $\ts'$ there
exists $\ts$, s.t., for any ball $\cU\subset
W^{\ts,2}\times W^{\ts,2} $ centered at the origin, there exists $\epsilon_*>0$ s.t.
\begin{equation}
  \label{stiH}
\sup_{0<\epsilon<\epsilon_*}\frac{\sup_{(\eta,\psi)\in\cU}\norma{J\nabla H_{WW}- J\nabla (H_0+\epsilon
  H_1+\epsilon^2H_2)}_{W^{\ts',2}\times W^{\ts',2}}}{\epsilon^3}<\infty
\ .
\end{equation}
\end{proposition}
The proof is postponed to Sect. \eqref{BNFill1}.


\subsection{Characteristic variables}\label{char.s}

We introduce the characteristic variables \eqref{rs}. Applying Remark
\ref{B} it is easy to see that the Hamilton equations take the form
\eqref{rsH}. 
Inserting in the various part of the Hamiltonian, one gets 
\begin{align}
  \label{h0rs}
H_0&=\int\frac{r^2+s^2}{2}dy\ ,
\\
\label{h1rs}
H_1&=\int_\R
\left(-\frac{1}{12}(r_y^2+s_y^2)+\frac{r^3+s^3}{4 }\right.
\\
\label{h1rs.1}
&\left. +\frac{r_ys_y}{6}-\frac{r^2s+rs^2}{4}
\right)dy
\\
\label{h2rs}
H_2&=\int\left(\frac{1}{2}\frac{r_{yy}^2+s_{yy}^2}{15}
-\frac{1}{4}(rr_y^2+ss_y^2)\right.
\\
\label{h2rs.1}
&\left. -\frac{1}{15}r_{yy}s_{yy}-\frac{1}{4}
(rs_y^2-2rr_ys_y+sr_y^2-2sr_ys_y)  
\right)dy
\end{align}
\begin{remark}
  \label{hj}
The Hamiltonian is the sum of terms, each of which is the integral
over $\R$ of a polynomial in $r,s$ and their derivatives. If a term is a
function of $r$ (and its derivatives) only, then it is invariant under
the flow of $H_0$ and thus it Poisson commutes with it, which means
that it is in normal form. The same is true if a term depends on $s$
and its derivatives only. 
\end{remark}

\begin{remark}
  \label{craig.1}
  If one restricts $H_0+\epsilon H_1$ to the manifold $s=0$ then one
  gets 
  \begin{align}
    \label{hcraig}
    H_{res}=\int
\left(\frac{r^2}{2}-\frac{1}{12}r_y^2+\frac{r^3}{4 }\right)dy\ ,
  \end{align}
  namely the Hamiltonian of a KdV equation in a reference frame translating with velocity
  $1$. 

  This is the procedure used by Craig and Groves in \cite{Cra} in
  order to deduce KdV as an equation describing the dynamics of water
  waves in this approximation.
\end{remark}

\section{Abstract Birkhoff normal form with no flow}\label{BNF}

\subsection{Birkhoff normal form in the finite dimensional
  case}\label{BNFfinite}

In this subsection we recall the algorithm of Birkhoff normal form in
the finite dimensional case. We will also present some explicit
formulae that will play a role in the water wave problem. 

We first introduce some notations. Let $\cP$ be a $2n$-dimensional
linear phase space endowed by a scalar product $\langle.;.\rangle$; we
denote by $J$ the Poisson tensor (namely a skewsymmetric invertible
linear opearator) and define the Hamiltonian vector field of a
Hamiltonian $G$ by $J\nabla G$. Furthermore, given a function $F$, we
denote by
$$\cL_{G}F:=dFJ\nabla G\equiv \langle \nabla F; J\nabla G\rangle  $$
its Lie derivative with respect to the vector field of $G$. In a
Hamiltonian framework this quantity is also called Poisson Brackets of
$F$ and $G$, and denoted by 
\begin{equation}
  \label{poi.bra}
\{F;G\}:=\cL_{G}H\ .
\end{equation}

Consider a family of Hamiltonian systems
\begin{equation}
  \label{H}
H(z,\epsilon)=\sum_{k\geq 0} \epsilon^k H_k(z)\ ,
\end{equation}
smooth in a neighborhood of the origin. In the following we will
not be interested in the size of the neighborhood, so we will not
specify the domain of functions, giving for understood that they are
smooth in a suitable neighborhood of the origin.

We are interested in the situation in which $H_0$ is a quadratic form
in $z$, whose Hamiltonian vector field generates a periodic flow. Then
it is well known that one can put the system in normal form at any
order. In particular the following version of Birkhoff normal form
theorem holds.

\begin{theorem}
\label{birkhoff} 
Fix an arbitrary positive integer $r\geq1$, then there exists a
canonical transformation $\Tr$ (defined in a neighborhood of the origin)
 which puts the system
(\ref{H}) in Normal Form at order $r$, namely such that
\begin{equation}
\label{eq:bir}
H\circ \Tr=H_0+\sum_{k=1}^{r}\epsilon^k Z_k+\O(\epsilon^{r+1})
\end{equation}
where $Z_k$ Poisson commutes with $H_0$, namely $
\left\{H_0;Z_k\right\}\equiv 0 $.
\end{theorem}

The idea of the proof is to construct iteratively a canonical transformation
putting the system in normal form. This means to first construct a
canonical transformation pushing the non normalized part of the
Hamiltonian to order $\epsilon^2$, then a transformation pushing it to
order $\epsilon^3$ and so on. Each of the transformations is constructed as the
flow of a suitable auxiliary Hamiltonian system (Lie
transform method).

We now perform explicitly the construction at
order three which is the one relevant for the water wave problem. 
\vskip10pt

Let $G $ be a smooth function, and consider the
corresponding Hamilton equations, namely
$$
\dot z=J\nabla G (z)\ ,
$$ 
denote by $\Phi_G^t$ the corresponding flow. 
\begin{definition}
\label{lie}
The map $\Phi_G^\epsilon$ will be called {\it Lie
  transform generated by $G $}.
\end{definition}

It is well known that
$\Phi_G^{\epsilon}$ is a canonical transformation.

We are now going to study the way a Hamiltonian changes when the
coordinate are subjected to a Lie transformation. Thus, let $F$ be a
smooth function and let $\Phi_G^\epsilon$ be the Lie transform generated
by a function $G $. To compute the
expansion of $F\circ\Phi_G^\epsilon$, first remark that
\begin{equation}
\label{dt}
\frac{d}{dt}F\circ\Phi^t_G=\left\{F ,G\right\}\circ \Phi_G^t
\end{equation}
so that, defining the sequence
\begin{equation}
\label{gl}
\alla F0 :=F\, ,\quad \alla Fl=\poisson{\alla F{l-1}}{G }\ ,\quad l\geq1\ ,
\end{equation}
one has $\forall r\geq0$
\begin{equation}
  \label{lietrans}
F\circ\Phi^\epsilon_{G}=\sum_{l=0}^r\frac{\epsilon^l}{l!}\alla
Fl+\O(\epsilon^{r+1}) \ .
\end{equation}

We come to the normalization procedure. We look for an auxiliary
Hamiltonian $G_1$ whose flow normalizes the Hamiltonian \eqref{H} at
first order. For a generic $G_1$, one has
\begin{align}
  \nonumber
  H\circ\Phi^\epsilon_{G_1}=(H_0+\epsilon H_1+\epsilon^2 H_2+\epsilon^3
H_3) \circ\Phi^\epsilon_{G_1}+ \O(\epsilon^{4})
\\
  \label{G.1}
=
H_0+\epsilon\poisson{H_0}{G_1}+\frac{\epsilon^2}{2}
\poisson{\poisson{H_0}{G_1}} 
{G_1} +\frac{\epsilon^3}{6}\poisson{\poisson{\poisson{H_0}{G_1}} 
{G_1}  }{G_1}  
\\
  \label{G.2}
+\epsilon H_1 +\epsilon^2\poisson{H_1}{G_1}+\frac{\epsilon^3}{2}
\poisson{\poisson{H_1}{G_1}} 
{G_1} 
\\
  \label{G.3}
+\epsilon^2 H_2 +\epsilon^3\poisson{H_2}{G_1}+\epsilon^3
H_3+\O(\epsilon^4) 
\end{align}
In order to determine $G_1$ in such a way that the terms of order
$\epsilon$ are in normal form, we recall the following well known
Lemma \cite{BG93}.

\begin{lemma}
  \label{solhomofinito}
Assume that the flow $\Phi_{H_0}^t$ is periodic of period $\tT$.  Define 
  \begin{equation}
    \label{z1.finito}
Z_1(z):=\frac{1}{\tT}\int_0^\tT H_1(\Phi^\tau(z))d\tau\ ,
  \end{equation}
  and $W_1:=H_1-Z_1$, then $Z_1$ is in normal form and
  \begin{equation}
    \label{g1.finito}
G_1(z):=\frac{1}{\tT}\int_0^\tT \tau W_1(\Phi^\tau(z))d\tau
  \end{equation}
  solves the \emph{homological equation}
  \begin{equation}
    \label{homo.finito}
    \poisson{H_0}{G_1}+W_1=0\ .
  \end{equation}
\end{lemma}
\proof Just compute
\begin{align*}
\poi{H_0}{G_1}(z)=-\frac{d}{dt}\big|_{t=0}G_1(\Phi^t_{H_0}(z))
=-\frac{1}{\tT}\int_0^{\tT}\tau
\frac{d}{dt}\big|_{t=0}W_1(\Phi^{t+\tau}_{H_0}(z))d\tau
\\
=-\frac{1}{\tT}\int_0^\tT\tau
\frac{d}{d\tau}W_1(\Phi^{\tau}_{H_0}(z))d\tau=-\frac{\tau W_1
  (\Phi^{\tau}_{H_0}(z))}{\tT}\big|_0^\tT
\\
+\frac{1}{\tT}\int_0^\tT
W_1(\Phi^{\tau}_{H_0}(z))d\tau=-W_1(z)\ .
\end{align*}
\qed

Using such a $G_1$, exploiting also \eqref{homo.finito} in order to
compute $\poi{H_0}{G_1}$, one gets
\begin{align}
  \nonumber
  H&\circ\Phi_{G_1}^\epsilon=H_0+\epsilon Z_1
  \\
  \label{H.2.1}
&+\epsilon^2 \left(\poi{Z_1}{G_1}+H_2+\frac{1}{2}\poi{W_1}{G_1}
\right)
\\
\label{H.2.2}
&+\epsilon^3\left( H_3 + \poi{H_2}{G_1}+\frac{1}{2}\poi{\poi{Z_1}{G_1}}
{G_1}+\frac{1}{3} \poi{\poi{W_1}{G_1}}
{G_1}  \right)
\\ \nonumber &+ \O(\epsilon^4)
\\
\nonumber
&=H_0+\epsilon Z_1+\epsilon^2 H_{2,1}+\epsilon^3 H_{3,1}+ \O(\epsilon^4)\ ,
\end{align}
where we denoted by $H_{2,1}$, resp. $H_{3,1}$ the brackets in
\eqref{H.2.1} resp. \eqref{H.2.2}.

Let $G_2$ be a further auxiliary Hamiltonian. One has
\begin{align*}
  H\circ\Phi_{G_1}^\epsilon\circ\Phi_{G_2}^{\epsilon^2}=H_0+\epsilon^2
  \poi{H_0}{G_2}+\epsilon Z_1 +\epsilon^3
  \poi{Z_1}{G_2}
  \\
  +\epsilon^2 H_{2,1}+\epsilon^3 H_{3,1} + \O(\epsilon^4)\ .
\end{align*}
Decomposing $H_{2,1}$ as in Lemma \ref{solhomofinito}, namely
\begin{equation}
  \label{h21}
H_{2,1}=Z_2+W_2\ 
\end{equation}
and determining $G_2$ as the solution of
\begin{equation}
  \label{homo.2.finito}
\poi{H_0}{G_2}+W_2=0\ ,
\end{equation}
one gets 
$$
 H\circ\Phi_{G_1}^\epsilon\circ\Phi_{G_2}^{\epsilon^2}=H_0+\epsilon
 Z_1+\epsilon^2 Z_2 +\epsilon^3 H_{3,2} +\O(\epsilon^4)\ ,
$$
 where, explicitly
 \begin{equation}
   \label{H32}
H_{3,2}=H_3 + \poi{H_2}{G_1}+\frac{1}{2}\poi{\poi{Z_1}{G_1}}
{G_1}+\frac{1}{3} \poi{\poi{W_1}{G_1}}
{G_1}+\poi{Z_1}{G_2} \ .
 \end{equation}
 To iterate a third time one has to decompose $H_{3,2}=Z_3+W_3$, to
 solve the homological equation
 \begin{equation}
   \label{homo.3}
\poi{H_0}{G_3}+W_3=0 \ ,
 \end{equation}
 and to transform using $\Phi_{G_3}^{\epsilon^3}$.

Of course one can iterate as many times as one wants. Here we
described the procedure at order 3, since in the case of the water
wave problem we do not have an abstract argument ensuring that $G_l$
belongs to a good class of objects and we need to compute it
explicitly. In particular, as we anticipated, at order $3$ we find
the first obstruction (see sect. \ref{NF}).

\subsection{Almost smooth maps}\label{BNFill1} We are now going to
generalize the above construction to the case where the vector field
of the function $G $ to be used to put the system in normal form does
not generate a flow. The idea is to approximate all the objects we
meet by their truncated expansion in $\epsilon$.

We will work in a scale of Banach spaces $\cB\equiv
\{\spazio {\ts}\}$. In the case of the water wave problem we will use the
space $\spazio {\ts}:=W^{\ts ,1}\times W^{\ts ,1} $ (since we will
work with the variables $(r,s)$). However it will be clear that everything
works in an abstract context. We will assume that for $\ts$
large enough the space $\spazio {\ts}$ is embedded in a Hilbert space,
whose scalar product $\langle .;.\rangle$ will be used to define the
gradient of functions.

In the case of the water wave problem the Hilbert space is
$L^2\times L^2$, so that the gradient will be with respect to the
standard $L^2\times L^2$ metric.

Furthermore we denote by $J$ a skewsymmetric operator
that we will use as the Poisson tensor. We assume that $\forall \ts$
there exists $\ts'$ such that $J:\ts'\to\ts$ is bounded.

In order to perform the proofs we will approximate the vector fields
by smooth objects. To this end we assume that there exists a sequence
of linear truncation operators $\{\Pi_N\}_{N\geq 0}$ which, for any
$\ts,\ts'$ are bounded as operators from $\spazio {\ts}$ to $\spazio {\ts
  '}$ and which converge to the identity as $N\to\infty$. Furthermore
we assume that $\Pi_N$ is self adjoint and commutes with $J$.

In the case of the water wave problem they are the standard truncations
in Fourier space. 

Following \cite{Bam13}, we will consider functions which have a weak
smoothness property. 

Let $\spazio\null\equiv\left\{\spazio {\ts}\right\}$ and
$\Kcs\equiv\left\{\Kcs^{\ts'}\right\}$ be two scales of Banach spaces, then
we give the following definition.

\begin{definition}
\label{def.smo.new}
A map $F$ will be said to be {\it almost smooth} if, $\forall
\tr,\ts'\geq 0$ there exist $\ts$ and an open neighborhood of the
origin $\cU_{\tr \ts \ts' }\subset\spazio {\ts}$ such that
\begin{equation}
\label{de.eq.1}
F\in C^{\tr}(\cU_{\tr \ts \ts' };\Kcs^{\ts'})\ .
\end{equation}

\end{definition}
We will use the same notation also when one of the two scales, or both,
is composed by a single space.

Furthermore, we will also deal with maps which depend on a small
parameter $\epsilon$. We will say that they are almost smooth if they
fulfill the above definition with the scale $\cB$ replaced by the
scale $\{\spazio {\ts}\times\R\}$, where $\R$ has been added as the
domain of $\epsilon$. \emph{In this case we will assume that the
  domain $ \cU_{\tr \ts \ts' }$ of \eqref{de.eq.1} has the form $
  \cU_{\tr \ts \ts' }=\cV_{\tr \ts \ts' }\times I_{\tr \ts \ts' } $
  with $\cV_{\tr \ts \ts' }\subset\spazio {\ts} $ and $I_{\tr \ts \ts'
  }$ an interval.  } The important point is that the size of the
open set $\cV_{\tr \ts \ts' }$ does not depend on $\epsilon$.

In the following the width of open sets does not play any role so
we will avoid to specify it. In particular we will often consider maps from
a Banach space  to some other
space, {\it by this we {\bf always} mean a map defined in an
  open neighborhood of the origin}. 

 We
remark that, according to the above definition, if $F$ is an almost
smooth map, then its differential has the property that
\begin{equation}
  \label{d.almost}
\forall l,r,\ \exists k_1,k_2\ ,\quad s.t.\quad dF(.)\in
C^{\tr}(\spazio {k_1};B(\spazio {k_2},\spazio l))\ . 
\end{equation}
In the following we will have to consider also the adjoint $dF(z)^*$
of $dF(z)$ with respect to the scalar product of the Hilbert space we
use for the computation of gradients. With a small abuse of
notation we will say that $dF^*$ is almost smooth if it has the
property \eqref{d.almost}.

\begin{definition}
  \label{ord}
In the rest of the paper we will write
  $$
A=B+\O(\epsilon^{r+1})
$$
if
$$
\frac{A-B}{\epsilon^{r+1}}
$$ is an almost smooth map.
\end{definition}

As a first application of this notion we give the proof of Proposition
\ref{approh}.

\noindent {\it Proof of Proposition \ref{approh}.} First we recall that,
in the original non scaled physical variables, the Hamilton equations
of \eqref{WW} are given by 
\begin{align}
  \label{ham.WW.1}
  \partial_t\eta&=G(\eta)\psi\ ,
  \\
  \label{ham.WW.2}
  \partial_t\psi&=-g\eta-\frac{1}{2}\psi_x^2+\frac{1}{2}
  \frac{\left(G(\eta)\psi+\eta_x\psi_x\right)^2}{1+\eta_x^2} \ .
\end{align}
After the scaling \eqref{scaling.1}, in particular the operator $G$ is
substituted by the scaled Dirichlet Neumann operator, that we now
denote by $G_{scal}$, studied in \cite{Lannes}, whose properties are
summarized in Proposition 3.44 of that book. In particular, by such a
proposition (and by Theorem 3.21 of \cite{Lannes}) one has that the map
$(\psi,\eta,\mu^2)\mapsto G_{scal}(\eta)\psi$ is almost smooth in the
scale $W^{\ts,2}\times W^{\ts,2}$.  It follows that the vector field
$J\nabla H_{WW}$ is almost smooth. Thus, from the formal computation
of Sect. \ref{scaling},
its truncated Taylor expansion in $\epsilon$ has the structure
$$
J\nabla H_{WW}=X_0 +\epsilon X_1+\epsilon^2 X_2+O(\epsilon^3)\ ,
$$
with $X_j=J\nabla H_j$, $j=0,1,2$, and this is the thesis. \qed

\begin{remark}
  \label{formal.rig}
From Proposition 3.44 of \cite{Lannes}, one has
$$
G_{scal}(\eta)\psi=
G^{(0)}\psi+G^{(1)}(\eta)\psi+G^{(2)}(\eta)\psi+O(\epsilon_1^3\mu^2)\ ,
$$
from which one gets a rigorous version of the formal formulae
\eqref{Gex}, \eqref{G1}. 
\end{remark}

\subsection{Lie transform with no flow}\label{BNFill}

Consider now an almost smooth vector field $X$ and define the sequence of
almost smooth vector fields
\begin{equation}
  \label{Xk}
\xalla0:=X\ ,\quad
  \xalla k:=d\xalla{k-1}X\ , \quad k\geq 1,
\end{equation}
\begin{remark}
  \label{smoo}
If $X$ is smooth
{\it as a map from $\spazio {\ts}$ to itself}, for some $\ts$, then denoting by
$\Phi^\epsilon$ the flow it generates, for any $r$ one has
\begin{equation}
  \label{flow}
\Phi^\epsilon(z)= z+\sum_{k\geq
  0}^r\frac{\epsilon^{k+1}}{(k+1)!}\xalla k(z)+\O(\epsilon^{r+1})\ ,
\end{equation}

This follows from the formula
$$
\frac{d^k}{dt^k}\left(X\circ\Phi^t\right)=d\xalla {k-1} \circ \Phi^t\ ,
$$
which is easily proven by induction. 
\end{remark}

{\bf Having fixed $X$ and $r\geq 1$,} we define
\begin{align}
  \label{T}
  T_{X,r,\epsilon}(z):=z+\sum_{k= 0}^{r-1}\frac{\epsilon^{k+1}}{(k+1)!}\xalla k(z)\ ,
  \\
  \label{tau}
\cT_{X,r,\epsilon} (z):=z+\sum_{k= 0}^{r-1}\frac{(-\epsilon)^{k+1}}{(k+1)!}\xalla k(z)\ .
\end{align}
In the following we will sistematically omit the indexes
${X,r,\epsilon}$ from $T$ and $\cT$.

We remark that both $T$ and $\cT$ are almost smooth maps Therefore
also $T\circ\cT$ and $\cT\circ T$ are almost smooth.

\begin{lemma}
  \label{appinv}
  One has
  \begin{equation}
    \label{appinv.1}
T\circ\cT=\uno+\O(\epsilon^{r+1})\ , \quad \cT\circ
T=\uno+\O(\epsilon^{r+1})\ .
  \end{equation}
\end{lemma}
\proof 
The proof is based on a regularization procedure. Using the
truncation operator $\Pi_N$ we define the truncated vector field by
\begin{equation}
  \label{trun}
X_N(z):=\Pi_NX(\Pi_Nz)\ .
\end{equation}
The flow it generates will be denoted by $\Phi_{N}^t$.

We consider $T\circ\cT$, the other case being equal.  Remark first
that such a quantity is smooth in $\epsilon$ so that it can be
expanded in Taylor series at any order. Thus the statement is
equivalent to the fact that the coefficient of order zero in the
expansion of $T\circ\cT$ is the identity, while the coefficients of
order from 1 to $r$ vanish. To prove this consider the sequence
$\xalla k_N$ generated by the vector field $X_N$ according to
\eqref{Xk}. Since $X_N$ is smooth (in the standard sense),
\eqref{flow} holds for it. Define the maps $T_{N}$ and
$\cT_{N}$ according to \eqref{T} and \eqref{tau} with $X_N$ in
place of $X$, then one has
$$
\Phi^\epsilon_{N}=T_{N}+\O(\epsilon^{r+1})\ ,\quad
\Phi^{-\epsilon}_{N}=\cT_{N}+\O(\epsilon^{r+1}) \ ,
$$
and
$$
\uno=\Phi^{\epsilon}_{N}\circ
\Phi^{-\epsilon}_{N}=T_{N}\circ\cT_{N}+
\left(\Tn\circ(\cTn+\O(\epsilon^{r+1}))-T_{N\circ}\cT_{N}
\right)+\O(\epsilon^{r+1}) \ ,
$$
from which
$$
\uno=\Tn\circ \cTn+\O(\epsilon^{r+1})\ .
$$
It follows that
\begin{equation}
  \label{resto1}
\frac{d^k}{d\epsilon^k}\big|_{\epsilon=0}\Tn\circ\cTn\equiv 0\ , \quad
\forall 1\leq k\leq r\ ,\quad \forall N\ .
\end{equation}
However, by construction $\Tn\to T$ and $\cTn\to \cT$ in
$C^{\tr}(\spazio {\ts '},\spazio {\ts})$ for all $\tr$ as $N\to\infty$, thus one gets 
\begin{equation}
  \label{resto1.1}
\frac{d^k}{d\epsilon^k}\big|_{\epsilon=0}T\circ\cT\equiv 0\ , \quad
\forall 1\leq k\leq r\ .
\end{equation}
which is the thesis. \qed

An immediate corollary of the above result is the following one.

\begin{corollary}
  \label{campiY}
  Let $Y$ and $X$ be almost smooth vector fields; fix $\ts'$, then
  there exist $\ts$, $\ts''$ and $\cU_{\ts\ts'}\subset\spazio {\ts}$ with the
  following property: let $\zeta
\in C^1([-\tT_0,\tT_0]; \cU_{\ts\ts'})$
 be a solution of
\begin{equation}
  \label{eq.Y}
\dot\zeta=d\cT(\zeta) Y(T(\zeta))\ , 
\end{equation}
then there exists $R \in C^1([-\tT_0,\tT_0];\spazio {\ts ''}) $ s.t.
$z(.):=T(\zeta(.))\in C^1([-\tT_0,\tT_0];\spazio {\ts '}) $ fulfills the
equation
\begin{equation}
  \label{quasi.Y}
\dot z=Y(z)+\epsilon^{r+1}R(t)\ .
\end{equation}
\end{corollary}
This is immediately seen by remarking that
$$
\dot z=\frac{d}{dt}T(\zeta(t))=dT(\zeta(t))\dot
\zeta=dT(\zeta)d\cT(\zeta) Y(T(\zeta))
=(\uno+\O(\epsilon^{r+1}))Y(T(\zeta))\ . 
$$

We come to the Hamiltonian case. Let $G \in C^{1}(\spazio {\ts},\R)$ and
$H\in C^{1}(\spazio {\ts},\R)$ be two Hamiltonian functions such that the
corresponding Hamiltonian vector fields $X:=J\nabla G $ and $J\nabla H$
are almost smooth.  Define the transformation $T$ according to
\eqref{T} and define the sequence $\alla Hl$ according to the recursive
definition \eqref{gl} and define
\begin{equation}
  \label{Htilde}
\widetilde H:=\sum_{l=0}^r\frac{\epsilon^l}{l!} \alla Hl\ ,
\end{equation}
then the main result of this section is the following Theorem
\begin{theorem}
  \label{teo.app}
Fix $\ts'$, then
  there exists $\ts$, $\ts''$ and $\cU_{\ts\ts'}\subset\spazio {\ts}$ with the
  following property: let $\zeta
\in C^1([-\tT_0,\tT_0]; \cU_{\ts\ts'})$, $0<\tT_0\leq\infty$, 
 be a solution of
  \begin{equation}
    \label{zeta}
  \dot \zeta=J\nabla \widetilde H(\zeta)\ ,
  \end{equation}
then there exists $R \in C^1([-\tT_0,\tT_0];\spazio {\ts '}) $ s.t.
$z(.):=T(\zeta(.))\in C^1([-\tT_0,\tT_0];\spazio {\ts ''}) $ fulfills the
equation
  \begin{equation}
    \label{zeta.1}
\dot z=J\nabla H(z)+\epsilon^{r+1}R(t)\ .
  \end{equation}
\end{theorem}

The rest of the section is devoted to the proof of this theorem. We
will proceed step by step.

Due to \eqref{nablecirc} and \eqref{tra.ham}, we have to study
$[dT(\zeta)]^*$, in particular in the case where $X=J\nabla G $.
First we remark that, in this case, for any $z\in \spazio {\ts}$,
$\ts$ sufficiently large, we have
\begin{equation}
  \label{self}
(d(\nabla G (z))^*=d\nabla G (z) \ ,
\end{equation}
where $(d\nabla G (z))^*$ is the adjoint with respect to the $L^2$
scalar product. Indeed, for $k\in\spazio {\ts}$, consider the
differential of the map $z\mapsto \langle k;\nabla G (z)\rangle$
applied to a vector $h$. We have
$$
d(\langle k;\nabla G  \rangle)h=\langle k;d(\nabla G ) h\rangle
=d(dG  h)k=d^2G (h,k)=d^2G (k,h)=  \langle h;d(\nabla G )
k\rangle\ ,
$$
which is the thesis.

Furthermore, since $J^*=-J$ is bounded, it follows that if $X=J\nabla G
$ is almost smooth, then also $(dX(z))^*=- d(\nabla G)J$ is almost
smooth.
\begin{lemma}
  \label{adj}
Let $X:=J\nabla G $ be an almost smooth vector field, then $(dT(z))^*$
is also almost smooth.
\end{lemma}
\proof We prove the result by induction on the vector fields
$\xalla k$. By the above remark the result is true for $\xalla 0$. By \eqref{Xk}
one has
$$
d\xalla k h= d^2\xalla {k-1}(X,h)+d\xalla {k-1}dX h\ .
$$
the adjoint of the second addendum is $dX^*[d\xalla {k-1}]^*$, so, by the
induction assumption it is almost smooth. Consider now the first addendum. 
The adjoint $L(z)$ of the linear operator $ d^2\xalla {k-1}(X,.) $ is
defined by
$$
\langle L(z)h_1;h_2\rangle=\langle d^2\xalla {k-1}(z)(X(z),h_2);h_1\rangle\ .
$$
Therefore one has to show that $\forall l$ $\exists k_1k_2$ s.t., if
$z\in \spazio {k_1}$, then $L(z)\in B(\spazio {k_1},\spazio {l})$ and furthermore the
dependence on $z$ is smooth. We start by fixing $z$, so that the
statement is equivalent to the existence of a constant $C$ s.t.
\begin{equation}
  \label{sti.dual}
\left|\langle L(z)h_1;h_2\rangle \right|\leq C
\norma{h_1}_{\spazio {k_2}}\norma{h_2}_{\spazio {-l}} \ .
\end{equation}
Actually, it is convenient to fix the argument of $X$ and to define
the operator $L_1(z)$ by
\begin{equation}
  \label{adj.2}
\langle L_1(z)h_1;h_2\rangle=\langle d^2\xalla {k-1}(z)(X(z_1),h_2);h_1\rangle\
\end{equation}
with fixed $z_1$ in a sufficiently smooth space. It is clear that, due
to the smooth dependence on $z_1$ it is sufficient to study the
operator $L_1$. Furthermore we denote simply $X(z_1)=X$ Now \eqref{adj.2} is
equal to
\begin{align*}
\langle d\left(d\xalla {k-1}(z)h_2\right)X;h_2  \rangle=d\left( \langle
d\xalla {k-1}(z)h_2;h_2  \rangle\right)X
\\
=d\left( \langle
h_2;[d\xalla {k-1}(z)]^*h_2  \rangle\right)X \ ,
\end{align*}
but, by the inductive assumption one has $[d\xalla {k-1}(.)]^*\in
C^{\tr}(\spazio {k_1},B(\spazio {k_2},\spazio l))$, therefore, if $X\in \spazio {k_1}$, which can
be ensured by taking $z_1$ smooth enough, the above quantity is
estimated by
$$
C\norma{X}_{\spazio {k_1}}\norma{ h_1}_{\spazio {k_2}}\norma{h_2}_{\spazio {-l}}\ ,
$$
which is the estimate that we had to prove. Smooth dependence on $z$
follows from the smooth dependence of $[d\xalla {k-1}(z)]^*$ on $z$. \qed

\begin{lemma}
  \label{trapoi}
Assume that $X=J\nabla G $ is an almost smooth vector field, then
one has
\begin{equation}
  \label{trapoi.1}
dT(\zeta)J[dT(\zeta)]^*=J+\O(\epsilon^{r+1})\ .
\end{equation}
\end{lemma}
\proof Let $G _N(\zeta):=G (\Pi_N\zeta)$ and denote
$X^{N}:=J\nabla G _N(\zeta)=\Pi_NJ(\nabla G )(\Pi_N\zeta)$. As
before, consider the corresponding flow
$\Phi_{N}^\epsilon=\Tn+\O(\epsilon^{r+1})$, which is a canonical
transformation. Thus one has
\begin{align}
\Pi_NJ\Pi_N=d \Phi_{N}^\epsilon(\zeta)\Pi_NJ\Pi_N [ d
  \Phi_{N}^\epsilon(\zeta) ]^*
=d\Tn(\zeta)\Pi_NJ\Pi_N[d\Tn(\zeta)]^*+\O(\epsilon^{r+1})\ .
\end{align}
It follows that, for all $N$ and for $1\leq l\leq r$, one has
\begin{align}
  \label{poi.0}
  d\Tn(\zeta)\Pi_NJ\Pi_N[d\Tn(\zeta)]^*\big|_{\epsilon=0}=\Pi_NJ\Pi_N\ ,
  \\
  \frac{d^l}{d\epsilon^l}\big|_{\epsilon=0}
  d\Tn(\zeta)\Pi_NJ\Pi_N[d\Tn(\zeta)]^*  =0 \ ,
\end{align}
but all these objects converge as almost smooth operators when
$N\to\infty$, and thus the thesis follows. \qed

\begin{corollary}
  \label{coro.ham.0}
  Let $\zeta(t)$ be a sufficiently smooth solution of
  \begin{equation}
    \label{lll}
\dot \zeta=J\nabla (H\circ T)(\zeta)\ ,
  \end{equation}
  then $z(t):=T(\zeta(t))$ fulfills
  \begin{equation}
    \label{rizeta}
\dot z=J\nabla H(z)+\epsilon^{r+1}R\ .
  \end{equation}
\end{corollary}
We are now ready for the proof of Theorem \ref{teo.app}. The main
point is that, from Corollary \ref{coro.ham.0}, in terms of the
variables $\zeta$, the system is Hamiltonian (up to a remainder of
order $\epsilon^{r+1}$) with Hamiltonian $H\circ T$. We now have have
the following Lemma

\begin{lemma}
  \label{resto..ham}
  One has
  \begin{equation}
    \label{re.fin}
    H\circ T=\sum_{l=0}^r\frac{\epsilon^l}{l!}\alla Hl+\epsilon^{r+1}R\ .
  \end{equation}
with $\alla Hl$ defined by \eqref{gl}, and $R$ having an almost smooth
vector field. 
\end{lemma}
\proof We start by showing that
$$
H^{(l)}=\frac{d^l}{d\epsilon^l}\big|_{\epsilon=0}H\circ T \ .
$$
which would show that $\epsilon^{r+1}R$ is the remainder of the Taylor
series of a smooth function (of $\epsilon$) and therefore $R$ is bounded uniformly with
respect to $\epsilon$. Consider the flow $\Phi^\epsilon_{N}$ of the
truncated vector field $X_N$, then one has
\begin{align*}
H\circ\Phi^\epsilon_{N}=H\circ (\Tn+\epsilon^{r+1}R)=H\circ \Tn+\left( H\circ
(\Tn+\epsilon^{r+1}R) - H\circ \Tn\right)
\\
= H\circ \Tn+\epsilon^{r+1}R
\ ,
\end{align*}
so that 
$$
\frac{d^l}{d\epsilon^l}\big|_{\epsilon=0}H\circ
\Tn=\frac{d^l}{d\epsilon^l}\big|_{\epsilon=0}H\circ
\Phi_{N}=H^{(l)}(\Pi_N. )\ ,
\quad \forall l\leq r \ . 
$$
Since this quantity converges to $H^{(l)}$ as $N$  tends to infinity, one
has the thesis. Reasoning in the same way on $\nabla(H\circ T)$, we
get
$$
\frac{d^l}{d\epsilon^l}\big|_{\epsilon=0}\left(\Pi_N\nabla
H(\Pi_N\Phi_{N} ^\epsilon)\right)= \Pi_N\nabla H^{(l)}(\Pi_N .)\ ,
$$
which, passing to the limit $N\to\infty$ shows that 
$$
\frac{d^l}{d\epsilon^l}\big|_{\epsilon=0}\nabla
(H\circ T)= \nabla H^{(l)}\ .
$$
Finally, by the almost smoothness of $\nabla (H\circ T)$, which
follows from eq. \eqref{nablecirc} and Lemma \ref{adj}, one has that
$\epsilon^{r+1}\nabla R$ is the remainder of a Taylor series of a
smooth function and thus the thesis follows. \qed

\noindent {\it Proof of Theorem \ref{teo.app}.}By Lemma
\ref{resto..ham} one has $\widetilde H=H\circ
T-\epsilon^{r+1} R_1$ with $R_1$ having an almost smooth vector field,
thus $\zeta(t)$ fulfills
$$
\dot \zeta(t)= J\nabla (H\circ T)-\epsilon^{r+1}J\nabla
R_1(\zeta(t))\ ,
$$
therefore, using \eqref{tra.ham}, we have
\begin{align*}
\dot z (t)=dT(\zeta(t))J\left[dT(\zeta(t))\right]^* \nabla
H(T(\zeta(t))) -\epsilon^{r+1}d T(\zeta(t)) J\nabla R_1(\zeta(t))
\\
= J\nabla H(z(t)) +\epsilon^{r+1} R_2(\zeta(t)) \nabla H(z(t))
-\epsilon^{r+1}d T(\zeta(t)) J\nabla R_1(\zeta(t))\ ,
\end{align*}
where we used Lemma \ref{trapoi}. But such an equation is the
thesis. \qed

\section{Hamiltonian Normal form for the water wave problem.}\label{NF}

In order to be able to apply the normal form procedure to the water
wave problem, we must be able to solve the Homological equation. This
is done with the help of a few lemmas. The first one is an abstract
lemma, the other two are really adapted to water wave problem. 

Consider the the homological equation
\begin{equation}
  \label{cohomo.abs}
\poi{H_0}G+ W=0\ .
\end{equation}

\begin{lemma}
  \label{sol.lemma.co}
  Assume that, for $\ts$ large enough, one has
  \begin{equation}
    \label{ker}
\lim_{\tau\to+\infty}(
W(\Phi_{H_0}^{-\tau}(z))+W(\Phi_{H_0}^{\tau}(z)))=0\ ,\quad \forall
z\in \spazio {\ts}\ ;
  \end{equation}
  if the following function $G $ is well defined, then it solves the
  homological equation \eqref{cohomo.abs}
  \begin{equation}
    \label{sol.cohomo.for}
G (z):=-\frac{1}{2}\int_{\R}\sgm(\tau)W(\Phi_{H_0}^{\tau}(z))d\tau\ . 
  \end{equation}
\end{lemma}
\proof Just compute 
\begin{align}
  \label{a.1}
\poi{H_0}G (z)=-\frac{d}{dt}\big|_{t=0}G (\Phi_{H_0}^{t}(z))
  \\
  =\frac{d}{dt}\big|_{t=0}
  \frac{1}{2}\int_{\R}\sgm(\tau)W(\Phi_{H_0}^{\tau+t}(z))d\tau
  \\
  =\frac{1}{2}\int_{\R}\sgm(\tau)\frac{d}{d\tau}W(\Phi_{H_0}^{\tau}(z))d\tau
  \\
  =-\frac{1}{2}\int_{-\infty}^0\frac{d}{d\tau}W(\Phi_{H_0}^{\tau}(z))d\tau
+\frac{1}{2}\int_0^{+\infty}\frac{d}{d\tau}W(\Phi_{H_0}^{\tau}(z))d\tau
\\
=-W( \Phi_{H_0}^{0}(z)) +\frac{W( \Phi_{H_0}^{-\infty}(z))+W(
  \Phi_{H_0}^{+\infty}(z))}{2} =-W(z)
\end{align}
\qed
 
Actually one can get an explicit formula for the solution of the
Homological equation. Before giving the result, we study a few
properties of the operator  $\partial^{-1}$ defined in
\eqref{dmeno.1.i}. First we remark that one also has
\begin{equation}
  \label{dmeno.1}
(\partial^{-1}u)(y)  =
\frac{1}{2}  \int_{\R}\sgm(y-y_1)u(y_1)dy_1 \ ,
\end{equation}
and that $\partial^{-1}:L^1\to L^{\infty}$ continuously. Then one has
$\partial(\partial^{-1}u)=u$. Furthermore, if $u$ is such
that $\lim_{\tau\to+\infty}(u(\tau)+u(-\tau))=0$ then one also has
$\partial^{-1}u_y=u$. We also remark that the property is automatic
for the functions of class $W^{2,1}$.

By the very definition of $\partial^{-1}$, its adjoint is
$-\partial^{-1}$.

Finally we introduce a notation which is very useful in order to shorten
the computations:

\emph{ In the following we denote
  \begin{equation}
    \label{notaz}
r_k:=\partial^kr\ ,\quad s_k:=\partial^ks\ ,\quad k\geq-1\ . 
  \end{equation}
}
We will consider functionals $W$ of the form
\begin{equation}
  \label{fw}
W(r,s)=\int_{\R}
P_1(r_{-1}(y),r(y),r_1(y),...,r_{n_1}(y))P_2(s_{-1}(y),s(y),s_1(y),...,s_{n_2}(y))
dy \ ,
\end{equation}
where $P_1:\R^{n_1+2}\to\R$ and $P_2:\R^{n_2+2}\to\R$ are polynomials. For brevity we will
simply denote
$$
P_1(r):=P_1(r_{-1}(y),r(y),r_1(y),...,r_{n_1}(y))\ .
$$
Sometimes we will denote
$$
P_1(r(y)):=P_1(r_{-1}(y),r(y),r_1(y),...,r_{n_1}(y))\ .
$$

We have the following Lemma
\begin{lemma}
  \label{sol.onde}
Assume that, $P_1(r)\in L^2(\R)$ whenever $r\in W^{\ts ,1}$, for
$\ts\gg1$, and similarly for $P_2(s)$. Then the solution
\eqref{sol.cohomo.for} of the homological equation \eqref{cohomo.abs}
with $W$ given by \eqref{fw} is given by
\begin{equation}
  \label{sol.app}
G(r,s):=-\frac{1}{2}\int_{\R}\left[\partial^{-1}P_1(r)\right] P_2(s)dy
\end{equation}
\end{lemma}

\proof We start by verifying that $W$ fulfills the
assumption \eqref{ker}. Fix some $K$, one has
\begin{align}
  \nonumber
W(\Phi^t_{H_0}(r,s))=
\int_{\R}P_1(r(y-t))P_2(s(y+t))dy=\int_{\R}P_1(r(y-2t))P_2(s(y))dy
\\
\label{azero}
=
\int_{-\infty}^KP_1(r(y-2t))P_2(s(y))dy
+\int_K^{+\infty}P_1(r(y-2t))P_2(s(y))dy \ . 
\end{align}
Consider first the first integral. It is estimated by
\begin{align*}
\left[ \int_{-\infty}^K|P_1(r(y-2t))|^2dy\right]^{1/2}\left[
  \int_{-\infty}^K|P_2(s(y))|^2dy\right]^{1/2}
\\
\leq \norma{P_2(s)}_{L^2}  \left[
  \int_{-\infty}^{K-2t}|P_1(r(y))|^2dy\right]^{1/2}\ , 
\end{align*}
but the last factor tends to zero when $t\to+\infty$, due to the
fact that $P_1(r)$ is square integrable. Treating the second integral
in \eqref{azero} in a similar way we get that $\lim_{t
  \to+\infty}W(\Phi^t_{H_0}(r,s))=0$. In a
similar way one gets $\lim_{t \to-\infty}W(\Phi^t_{H_0}(r,s))=0$.

We now use the formula \eqref{sol.cohomo.for} to compute $G$.  Making
the change of variables $$y_1=y-\tau\ ,\quad y_2=y+\tau\ ,
$$ one has
\begin{align}
G =-\frac{1}{2}\int_{\R}d\tau \sgm(\tau)\int_{\R}dy
P_1(r(y-\tau))P_2(s(y+\tau))
\\
= \frac{1}{2}(-)\frac{1}{2}\int_{\R^2}
\sgm(y_2-y_1) P_1(r(y_1))P_2(s(y_2))dy_1dy_2
\\
=-\frac{1}{2}\int_{\R}dy_2
[(\partial^{-1}P_1(r))(y_2)]P_2(s(y_2))\ .
\end{align}
\qed

Actually we do not have an abstract theorem ensuring that the
Hamiltonian vector field of $G$ is an almost smooth map. We now
compute explicitly the second order normal form and compute the
structure of the first two generating functions in order to show that
their vector field is almost smooth. Furthermore we compute some
terms of $G_3$ in order to show that the corresponding vector field is
not well defined, so that we cannot perform (at least with this
algorithm) a third step completely eliminating the interaction between
right going waves and left going waves.

In order to simplify the notation and the computation,
given a
functional which is of the form
\begin{equation}
  \label{not.1}
W(r,s)=\int_{\R}w(r(y),s(y))dy\ ,
\end{equation}
with $w(r,s)=P_1(r)P_2(s)$, we will always denote by lower case letter
the density which is integrated to get the functional denoted with the
corresponding capital letter.

\begin{remark}
  \label{gradient}
Given a functional $W$ as in \eqref{not.1}, the corresponding gradient
is given by
\begin{equation}
  \label{grad}
\nabla _r W(r,s)=\sum_{k\geq-1}^{n_1}(-\partial^k)\frac{\partial
  w}{\partial r_k}
\end{equation}
and similarly for the gradient with respect to the $s$ variable.
\end{remark}

\begin{lemma}
  \label{local}
Assume that $W$ is of the form \eqref{fw} with $P_1$ and $P_2$
fulfilling the assumptions of Lemma \ref{sol.onde}. Assume also that
$P_1$ and $P_2$ are monomyals that do not depend on $r_{-1}$ and
$s_{-1}$ respectively. Then the solution $G$ of the homological
equation \eqref{cohomo.abs} defined by \eqref{sol.app} has an almost
smooth vector field.
\end{lemma}
\proof Up to the factor $1/2$ and exploiting the skew symmetry of
$\partial^{-1}$, one has $g=P_1(r)\partial^{-1}P_2(s)$, from which
\begin{align*}
-\partial \nabla_r G=-\partial \sum_{k\geq 0}^{n_1}(-\partial)^k\left(\frac{\partial
  P_1  }{\partial r_k} \partial^{-1}P_2  \right)
\\
= (\partial^{-1} P_2(s))
\sum_{k\geq 0}^{n_1}\left((-\partial)^{k+1}\frac{\partial
P_1  }{\partial r_k}   \right) +\text{{\rm local\ terms}}
\end{align*}
where, by local terms, we mean terms not involving $\partial^{-1}$. 

We prove now that $\forall k\geq0$, $(-\partial)^{k+1}\frac{\partial
  P_1 }{\partial r_k}\in W^{\ts ,1}$. Indeed, if
$\frac{\partial P_1 }{\partial r_k}$ is not a constant,
then the result follows from the algebra property of $W^{\ts ,1}$,
while, if it is a constant, then $\partial^{k+1}$ annihilates
it. Thus, since the product of a function of class $L^1$ and a
function of class $L^\infty$ is still of class $L^1$ the result
follows for the $r$ component. Similarly one gets the result for the
$s$ component.\qed
\unp

We now proceed in the explicit computation of $z_i,w_i$ and $g_i$.

Consider $H_1$ as given \eqref{h1rs}, \eqref{h1rs.1}, in which
$Z_1=\eqref{h1rs} $ and  $W_1=\eqref{h1rs.1} $, so that one has 
\begin{equation}
  \label{z.1}
z_1=-\frac{1}{12}(r_1^2+s_1^2)+\frac{r^3+s^3}{4 }\ ,\quad
w_1 =
\frac{r_1s_1}{6}-\frac{r^2s+rs^2}{4}\ .
\end{equation}
From this, by Lemma \ref{sol.onde} and the skewsymmetry of
$\partial^{-1}$,
\begin{equation}
  \label{g.1}
g_1=\frac{r_1s}{12}-\frac{r^2s_{-1}-r_{-1}s^2}{8}\ ,
\end{equation}
In particular, by Lemma \ref{local} we know that its vector field is
almost smooth.

Furthermore, one has
\begin{align}
  \label{campow1r}
  \nabla_rW_1&=-\frac{1}{6}s_2-\frac{rs}{2}-\frac{s^2}{4} 
\\
  \label{campow1s}
  \nabla_sW_1&=-\frac{1}{6}r_2-\frac{rs}{2}-\frac{r^2}{4} 
\\
\label{campog1r}
  \nabla_rG_1&=-\frac{rs_{-1}}{4}-\partial^{-1}\frac{s^2}{8}
-\frac{s_1}{12}\ , 
\\
\label{campog1s}
  \nabla_sG_1&=\frac{sr_{-1}}{4}+\partial^{-1}\frac{r^2}{8}
+\frac{r_1}{12}\ .
\end{align}
So, in particular the vector field of $G_1$ is
\begin{align}
  \label{camp.ham.g1}
(r-\text{{\rm
      component}})=\frac{r_1s_{-1}+rs}{4}+\frac{s^2}{8}+
  \frac{s_2}{12} \\ (s-\text{{\rm
      component}})=\frac{s_1r_{-1}+rs}{4}+\frac{r^2}{8}+\frac{r_2}{12}\ .
\end{align}
\begin{remark}
  \label{calore}
If in the expressions of the vector field of $G_1$ we neglect the
nonlinear terms, the corresponding equations of motion turn out to be
\begin{equation}
  \label{calore.1}
\left\{  \begin{matrix}
  \dot r=\frac{s_2}{12}
  \\
  \dot s=\frac{r_2}{12}
\end{matrix}\right. \Longrightarrow \left\{  \begin{matrix}
  \ddot r=\frac{r_4}{144}
  \\
  \ddot s=\frac{s_4}{144}
\end{matrix}\right.
\end{equation}
which is clearly ill posed. It follows in particular that the problem
of existence and uniqueness for the Hamilton equations of $G_1$ is a
nontrivial one. With our approach we do not need to study it.
\end{remark}

We now compute $H_{2,1}=W_2+Z_2$ (cf. \eqref{H.2.1}), in particular we
will get an explicit expression for the
terms contributing to $Z_2$. For the terms contributing to $W_2$, we
will neglect the precise value of the coefficients of the various
terms, that will be conventionally put equal to 1.

First remark that $\poi{Z_1}{G_1}$ does not contribute to $Z_2$, so we
will only compute its general structure.

To start with compute (with a small abuse of notation) 
\begin{align*}
\poi{W_1}{G_1}=\langle \nabla_s W_1;\partial \nabla_s G_1\rangle-
\langle \nabla_r W_1;\partial \nabla_r G_1\rangle
\\
=\left(-\frac{1}{6}r_2-\frac{rs}{2 }-\frac{r^2}{4 }\right)
\left(\frac{s_1r_{-1}+rs}{4 }+\frac{r^2}{8 }+\frac{r_2}{12}\right)
\\
+
\left(-\frac{1}{6}s_2-\frac{rs}{2 }-\frac{s^2}{4 }\right) 
\left(\frac{r_1s_{-1}+rs}{4 }+\frac{s^2}{8 }+\frac{s_2}{12}\right)
\\
=-\frac{1}{24 }r^2r_{2}-\frac{1}{72}r_2^2
-\frac{1}{32}r^4
-\frac{1}{24 }s^2s_{2}-\frac{1}{72}s_2^2
-\frac{1}{32}s^4
\\
+r_2s_1r_{-1}+r_2rs+rss_1r_{-1}+r^2s^2+sr^3+r^2s_1r_{-1}+s_2r_1s_{-1}
\\
+s_2rs+rsr_1s_{-1} +rs^3+rss_2+s^2r_1s_{-1}+rs^3
\end{align*}
while we have 
\begin{align*}
\poi{Z_1}{G_1}=(s_2+s^2)  (s_1r_{-1}+rs+{r^2}+{r_2} ) +  (r_2+r^2)(
r_1s_{-1}+rs+{s^2}+{s_2}  )
\\
=s_2s_1r_{-1}+s_2r^2+s_2r_2+s^2r^2+s^2r_2+r_2r_1s_{-1}+r_2s^2+r_2r_1s_{-1}
\\
+ \text{{\rm terms\ already\ contained \ in\ }}\poi{W_1}{G_1}
\end{align*}
Integrating by parts the first term of $\frac{1}{2}\poi{W_1}{G_1}$ and
adding the terms coming from
$H_2$, we have
\begin{align}
  \label{z.2}
z_2=-\frac{5 }{24 }rr_{1}^2+\frac{19}{720}r_2^2
-\frac{1}{64}r^4 
\\
-\frac{5 }{24 }ss_{1}^2+\frac{19}{720}s_2^2
-\frac{1}{64}s^4 \ ,
\end{align}
so that, its gradient and the corresponding vector field are given by
\begin{align}
  \label{z2campo}
\nabla_rZ_2=\frac{5}{24}r_1^2+\frac{5}{12}
 rr_2+\frac{19}{360}r_4-\frac{1}{16}
r^3
\\
\label{z2campr}
-\partial\nabla_rZ_2=
-\frac{5}{6}r_1r_2-\frac{5}{12}
rr_3
- \frac{19}{360}r_5+\frac{3}{16}
r^2r_1\ .
  \end{align}
Concerning $W_2$, one has 
\begin{align}
  \label{w2}
w_2=r_2s_1r_{-1}+r_2rs+rss_1r_{-1}+r^2s^2+sr^3+r^2s_1r_{-1}+s_2r_1s_{-1}
\\
+s_2rs+rsr_1s_{-1} +rs^3+rss_2+s^2r_1s_{-1}+rs^3
\\
\label{w2.1}
+s_2s_1r_{-1}+s_2r^2+s_2r_2+s^2r^2+s^2r_2+r_2r_1s_{-1}+r_2s^2\ .
\end{align}
\begin{lemma}
  \label{g2} Let $G_2$ be given by \eqref{sol.app} with $P_1$ and
  $P_2$ given by the different terms of \eqref{w2}-\eqref{w2.1}.
Then the vector field of $G_2$ is almost smooth.
\end{lemma}
\proof According to Lemma \ref{local}, we only have to check the terms
coming from nonlocal terms in $w_2$, namely
\begin{align*}
w_2^{nl}=r_2s_1r_{-1}+rss_1r_{-1}+r^2s_1r_{-1}+s_2r_1s_{-1}
+rsr_1s_{-1} +s^2r_1s_{-1}
\\
+s_2s_1r_{-1}+r_2r_1s_{-1}
\\
= r_2r_{-1}s_1+rr_{-1}\partial(s^2)+r^2r_{-1}s_1+r_1s_2s_{-1}
+\partial(r^2) ss_{-1} +r_1s_{-1} s^2
\\
+r_{-1}\partial(s_1^2)+\partial(r_1^2)s_{-1}\ , 
\end{align*}
from which
\begin{align*}
g_2^{nl}= r_2r_{-1}s+rr_{-1}s^2+r^2r_{-1}s+rs_2s_{-1}
+r^2 ss_{-1} +rs_{-1} s^2
+r_{-1}s_1^2+r_1^2s_{-1}\ . 
\end{align*}
By the same argument as in the proof of Lemma \ref{local}, the vector
field corresponding to each term of the above equation has an almost
smooth vector field. \qed

As a consequence one can use $G_2$ to put the system in normal form at
order $\epsilon^2$. To give a precise statement consider the
 Hamiltonian
\begin{align}
  \label{hanz}
  H_{Z}(r,s):=H_0(r,s)+\epsilon Z_1(r,s)+\epsilon^2Z_2(r,s)\ ,
\end{align}
with $Z_1$ given by \eqref{z.1} and $Z_2$ by \eqref{z.2}.

\begin{theorem}
  \label{main.ham}
For any $\ts'$ there exists $\epsilon_*>0$ and $\ts$, $\ts''$, s.t., if
$0<\epsilon<\epsilon_*$, then there exists a map $T_H:B_1^{\ts}\to \spazio
{\ts''}$, with the following properties
\begin{itemize}
\item[(i)] $T_H(r,s)-(r,s)$ is a polynomial in
  $r_k,s_k$, $k=-1,...,5$,
\item[(ii)] $\sup_{(r,s)\in
  B_1^{\ts}}\norma{T_H(r,s)-(r,s)}_{\spazio{\ts''}} \leq \epsilon $,
\item[(iii)] Let $I_{\epsilon}$ be an interval containing 0 and let
  $z(.)=(r(.),s(.))\in C^1(I_{\epsilon};B_1^{\ts})$ be a solution of the
  Hamiltonian system \eqref{hanz} define
  \begin{equation}
  \label{trans}
z_h\equiv(r_h,s_h):=T_{H}(r,s)\ .
\end{equation}
Then there exists $R\in C^1(I_{\epsilon},W^{\ts',2}\times
W^{\ts',2})$ s.t. one has
  \begin{equation}
    \label{main.diff}
\dot z_h(t)=J\nabla H_{WW}(z_h(t)) + \epsilon^3 R(t)\ ,\quad \forall t\in
I_\epsilon\ ,
  \end{equation}
  where $H_{WW}$ is the Hamiltonian \eqref{H.riscl.i} of the water wave problem
  rewritten in the variables $(r,s)$.
\end{itemize}
\end{theorem}

\proof Define $X_1:=J\nabla G_1$ with $G_1$ given by \eqref{g.1} and
\begin{equation}
  \label{t1.eq}
T_1(z):=z+\epsilon X_1(z)+\epsilon^2 dX_1(z)X_1(z)\ ,
\end{equation}
define also
\begin{equation}
  \label{t2.eq}
T_2(z)=z+J\nabla G_2(z)\ ,
\end{equation}
with $G_2$ as described in the statement of Lemma \ref{g2}. Define
$T_H:=T_1\circ T_2$, then Theorem \ref{teo.app} shows that there
exists $R_h\in C^1(I_{\epsilon},W^{\ts'+1,1}\times
W^{\ts'+1,1})$ s.t.
$$
\dot z_{h}=J\nabla (H_0+\epsilon H_1+\epsilon H_2)+ \epsilon^3 R_h \ .
$$
Adding the remainder coming from the truncation of the Hamiltonian
(c.f. Proposition \ref{approh}) and exploiting the embedding $W^{\ts'+1,1}\times
W^{\ts'+1,1}\subset W^{\ts',2}\times
W^{\ts',2}$ one gets the result. \qed

Then one would like to make at least a third step. As we anticipated
there are obstructions that we now describe.
Using the formula \eqref{H32}, one sees that $W_3$ contains in
particular the term $\{Z_2,G_1\}$. Thus in particular it contains a
monomyal coming from the terms $r_2^2$ in $Z_2$ and the term
$r^2s_{-1}$ in $g_1$. This gives rise to a nonlocal term in $W_3$ which is
$$
w_3^{bad}:= r_4r_1s_{-1}\ ,
$$
which in turn give rise to
$$g_3^{bad}= \partial^{-1}(r_4r_1)s_{-1}\ ,
$$
whose integral over $\R$ is, in general infinite. Even working
formally, one can compute the corresponding term in the Hamiltonian
vector field. It is given by
$$
\partial_4(r_1s_{-2})+ \partial(r_4s_{-2})=r_5 s_{-2}  +\text{{\rm
    local\ terms}}\ ,
$$
which is not well defined, since the operator $\partial^{-2}$ is in
general not defined on $W^{\ts ,1}$.

Actually this argument is not conclusive, since there could be terms
compensating $w_3^{bad}$ or additive terms which transform such a term
in something of the form $\partial^3(r_1)s_{-1}$, which would give
rise to well behaved terms. However the verification of this requires
much longer computations that we leave for future work.

\section{Kodama's theory}\label{kodama}

Given a Hamiltonian system of the form
\begin{equation}
  \label{kod.1}
K_0(s)+\epsilon K_1(s)+\epsilon^2Z_2(s)\ ,
\end{equation}
with $K_0$  and $K_1$ given by \eqref{kdv0} and \eqref{kdv1} and
\begin{equation}
  \label{kod.1.1}
Z_2(s)=\int_{\R}\left(b_1ss_1^2+b_2s_2^2+b_3s^4\right)dy\ ,\quad b_j\in\R
\end{equation}
Kodama \cite{Kod85,Kod87,Kod87a} (but we make here reference to the
review paper \cite{HK09}) has shown that there exists a coordinate
transformation of the form
\begin{align}
  \label{kod.2}
s&=T_K(u):=u+\epsilon\cX(u)+\epsilon^2d\cX(u)\cX(u)\ ,
\\
\label{kod.3}
\cX(u)&:=a_1u^2+a_2u_2+a_3u_1u_{-1}\ ,\quad a_j\in\R
\end{align}
which conjugates the Hamilton equations of \eqref{kod.1} to the
Hamilton equations of 
\begin{equation}
  \label{kod.4}
 K_0(u)+\epsilon K_1(u)+\epsilon^2 c_2K_2(u)\ ,
\end{equation}
with $K_2$ given by \eqref{kdv2} and a suitable $c_2$ to be
determined.

So the result of Theorem \ref{main.0} directly follows from Kodama's
theory.

\vskip10pt
For the sake of completness we are now going to summarize such a
theory.

First consider the Hamilton equations of \eqref{kod.1}, which have the
form
\begin{equation}
  \label{kod.5}
\dot s =\cY_0(s)+\epsilon\cY_1(s)+\epsilon^2\cY_2(s)\ ,
\end{equation}
with $\cY_0(s)=\partial_y\nabla K_0(s)$ and so on. Then remark that
\eqref{kod.2} is just the second order expansion of the time
$\epsilon$ flow of the auxiliary equation $\dot u=\cX(u)$. Thus, by
repeating at a non Hamiltonian level the computations of
sect. \ref{BNFill}, one gets that the equations fulfilled by $u$ (as
defined by \eqref{kod.2}) are
\begin{equation}
  \label{kod.44}
\dot u=\widetilde \cY(u)\ ,
\end{equation}
with
\begin{align*}
\widetilde \cY=\cY_0+\epsilon
           [\cY_0;\cX]+\frac{\epsilon^2}{2}\left[\left[\cY_0;\cX\right];\cX
             \right]+\epsilon\cY_1+\epsilon^2\left[\cY_1;\cX\right]
           +\epsilon^2\cY_2  +O(\epsilon^3)
\\
=\cY_0+\epsilon\cY_1+\epsilon^2\left(\cY_2+\left[\cY_1;\cX\right]\right)
+O(\epsilon^3)  \ ,
\end{align*}
where we denoted
$$
\left[\cY;\cX\right](u):=d\cY(u)\cX(u)-d\cX(u)\cY(u)\ ,
$$
and used the fact that $[\cY_0;\cX]=0$, since $\cY_0$ is the generator
of the translations (and also follows by direct computation). 

Thus, one has to look for the values of the constants $a_j$ in
\eqref{kod.3} such that
\begin{equation}
  \label{kod.33}
\cY_2+\left[\cY_1;\cX\right]=c_2\partial \nabla K_2\equiv
c_2(u_5+5u_3u+10 u_2u_1+\frac{35}{8}u_1u^2)\ . 
\end{equation}

Now, a long, but straightforward computation (which might be correct)
shows that
\begin{align*}
\cY_2+\left[\cY_1;\cX\right]=\left(\frac{1}{3}a_2+2b_2\right)u_5+
\left(\frac{1}{3}a_1+\frac{1}{2}a_3-2b_1\right) u_3u
\\
+\left(a_1-2a_2+\frac{5}{6}a_3-4b_1\right)u_2u_1+
\left(\frac{7}{2}a_1+\frac{3}{4} a_3+7b_3\right)u_1u^2 \ .
\end{align*}
this leads to impose the system for the unknowns $(a_1,a_2,a_3,c_2)$
\begin{align*}
  c_2&=\frac{1}{3}a_2+2b_2
  \\
  5c_2&=\frac{1}{3}a_1+\frac{1}{2}a_3-2b_1
  \\
  10c_2&=a_1-2a_2+\frac{5}{6}a_3-4b_1
  \\
  \frac{35}{8}c_2&=\frac{7}{2}a_1+\frac{3}{4} a_3+7b_3
\end{align*}
which can be solved explicitly, giving in particular
\begin{equation}
  \label{c2}
c_2=\left(7b_3+3b_1+81b_2\right)\frac{8}{389}=\frac{299}{389}\ ,
\end{equation}
where the last equality is obtained by inserting the values of $b_j$
coming from \eqref{z.2}. 

To conclude the proof one has just to define the transformation
\begin{equation}
  \label{tepsilon}
T_{\epsilon}:=T_H\circ T_K
\end{equation}
and remark that it still has the property (i) of Theorem
\ref{main.0}.

\addcontentsline{toc}{chapter}{Bibliography}



\end{document}